\documentclass[final,3p]{elsarticle}

\usepackage{amsmath}
\usepackage{amssymb}
\usepackage{siunitx}

\newcounter{bla}

\journal{Computer Physics Communications}
\usepackage{graphicx}
\usepackage{bm}

\usepackage{hyperref}
\usepackage[capitalize]{cleveref}
\usepackage{csquotes}
\usepackage{dsfont}
\usepackage{booktabs}
\usepackage{listings}
\usepackage{xcolor}
\usepackage{makecell}

\renewcommand*{\phi}{\varphi}

\let\Re\relax
\DeclareMathOperator{\Re}{Re}

\crefname{lstlisting}{Listing}{Listings}
\Crefname{lstlisting}{Listing}{Listings}

\usepackage{comment}

\usepackage{hyperref}
\usepackage{xcolor}

\definecolor{linkcolor}{HTML}{0176ba}
\definecolor{urlcolor}{HTML}{0176ba} 
\definecolor{citecolor}{HTML}{900020}
\hypersetup{pdfstartview=FitH,  linkcolor=linkcolor,urlcolor=urlcolor, citecolor=citecolor, colorlinks=true}

\usepackage{orcidlink}

\begin{document}

\begin{frontmatter}



\title{\textit{acoustotreams} -- A Python package for acoustic-wave scattering based on the \textit{T}-matrix method}


\author[a]{Nikita Ustimenko\corref{author}\orcidlink{https://orcid.org/0000-0002-5137-493X}}
\author[a,b,c]{Carsten Rockstuhl\orcidlink{https://orcid.org/0000-0002-5868-0526}}

\cortext[author] {Corresponding author.\\\textit{E-mail address:} nikita.ustimenko@kit.edu}
\address[a]{Institute of Theoretical Solid State Physics, Karlsruhe Institute of Technology, Kaiserstrasse 12, D-76131, Karlsruhe, Germany}
\address[b]{Institute of Nanotechnology, Karlsruhe Institute of Technology, Kaiserstrasse 12, D-76131, Karlsruhe, Germany}
\address[c]{Center for Integrated Quantum Science and Technology (IQST), Karlsruhe Institute of Technology, Wolfgang-Gaede-Str. 1, 76131 Karlsruhe, Germany}

\begin{abstract}
The transition-matrix (\textit{T}-matrix) method has established itself as a prominent technique for computing the scattering response from spatially localized objects. The suitability becomes apparent particularly when considering not just isolated objects but also large ensembles of aperiodically or even periodically arranged objects. A versatile implementation of the method is provided by the \textit{treams} program, which efficiently computes the electromagnetic response of scatterers in various arrangements~\cite{1}. 
Here, we rely on this framework and present a new program, \textit{acoustotreams}, dedicated to simulating the acoustic scattering of pressure waves by clusters of particles, both with and without periodic boundary conditions. The computations are performed using the \textit{T}-matrix method with scalar spherical and cylindrical waves as basis sets, and the scattering matrix (\textit{S}-matrix) method in the basis of scalar plane waves for stratified media.
The underlying theory is presented alongside the program structure and illustrative examples. The code is open-source and available on the Python Package Index for Linux, Windows, and macOS. Version control is maintained through GitHub, where we also provide automated tests, documentation, and detailed examples. We expect this work to contribute to the field of numerical methods for multiple-scattering problems by offering a computational framework capable of a comprehensive description of pressure-acoustic scattering in artificial media, including well-established metamaterials and metasurfaces.    







\noindent \textbf{PROGRAM SUMMARY}

\begin{small}
\noindent
{\em Program Title:} acoustotreams                                          \\
{\em Developer's repository link:} \href{https://github.com/NikUstimenko/acoustotreams}{https://github.com/NikUstimenko/acoustotreams} \\
{\em Licensing provisions:} MIT \\
{\em Programming language:} Python 3                             \\
{\em Nature of problem:} Simulation of the scattering of acoustic waves at arbitrarily shaped objects and arrangements thereof. The possible arrangements can be clusters of a large number of (possibly) different scatterers with an arbitrary rotation and position, or periodic arrangements of an identical unit cell containing one or multiple scatterers along multiple dimensions. Studying such problems with conventional full-wave solvers becomes prohibitively slow -- or even infeasible -- for large numbers of scatterers, especially when multiple system parameters must be swept. This challenge can be addressed using the \textit{T}-matrix method, which should be implemented conveniently and comprehensively. \\
{\em Solution method:} The scattering of an arbitrary acoustic wave by a given scatterer is fully described by its acoustic \textit{T}-matrix, which is independent of the incident field. Here, the incident and scattered fields are expanded in basis functions, and the amplitudes of the basis functions are assembled in a vector. The scattering problem is then solved as a matrix-vector product. Within the \textit{T}-matrix framework, acoustic interactions between multiple scatterers are expressed by analytically known translation coefficients in the absence of periodic boundary conditions, or by lattice sums when periodic boundary conditions are present. Rapid convergence of lattice sums is achieved using the Ewald summation method. Depending on the specific setting, scalar spherical, cylindrical, or plane waves can be used as basis functions. For stratified media, computations are carried out using \textit{S}-matrices. For convenience, all functionalities are accessible through class instances that track physical information encoded in annotations, such as the basis, the wavenumber, etc. \\
{\em Additional comments including restrictions and unusual features:} Background media cannot sustain shear waves, and their material parameters (mass density and speed of sound) must be real-valued scalars. The circumscribing spheres of scatterers should neither overlap nor touch. Scatterers that do not meet these conditions are not recommended in multiple-scattering computations. \\
   \\

\end{small}
   \end{abstract}
\end{frontmatter}

\section{Introduction}

Suitably shaped acoustic scatterers and their spatially ordered assemblies, such as metamaterials, metasurfaces, phononic crystals, or clusters comprising large numbers of arbitrarily shaped scatterers, offer numerous opportunities for the on-demand control of sound-matter interactions. Demonstrated applications include antennas for directional scattering~\cite{Morvan2014Dec,Wei2020Aug,Timankova2026Jan}, ultrasound modulators~\cite{Ma2020Sep}, acoustic metalenses~\cite{Mei2023Jul}, noise insulation systems~\cite{Krasikova2023Jan}, acoustic waveguides~\cite{Sainidou2006May,Wang2024May}, acoustic tweezers~\cite{Marzo2015Oct}, and high-$Q$ resonators enabled by bound states in the continuum~\cite{He2025Jul}.

Acoustic scattering can be modeled using various numerical methods, including the boundary element method~\cite{Kirkup2019Apr}, finite element method (FEM)~\cite{Everstine1997Nov,Wu2015}, finite-difference time-domain method~\cite{Treeby2010Mar}, and null-field boundary integral equation method~\cite{Chen2010Aug}. However, the design and optimization of metamaterials require extensive simulations, often involving parameter sweeps and a deep understanding of the underlying physical mechanisms. This multiple-scattering problem can be efficiently addressed using the transition matrix (\textit{T}-matrix) method--a generalization of Mie theory~\cite{Bohren1998Apr,Mishchenko2009Sep} for arbitrarily shaped electromagnetic~\cite{Waterman1965Aug} and acoustic~\cite{Waterman1969Jun,Waterman2009Jan} scatterers, originally introduced by Waterman. The method was later extended to systems with an arbitrary number of optical~\cite{Peterson1973Nov} and acoustic~\cite{Peterson1974Sep} scatterers by Peterson and Ström. 

Recently, the \textit{T}-matrix-based code \textit{treams} was published~\cite{Beutel2024Apr}. This open-source software combines a highly accessible object-oriented interface with efficient computational approaches, forming a comprehensive framework for describing electromagnetic scattering in both finite clusters and periodic lattices. For periodic systems, performance critically depends on evaluating lattice sums, which typically converge slowly. In \textit{treams}, this issue is addressed using the Ewald summation technique~\cite{Beutel2023Jan}, enabling efficient computation for arbitrary one-dimensional (1D), two-dimensional (2D), and three-dimensional (3D) lattices with complex unit cells, i.e., unit cells containing multiple (possibly different) scatterers. Furthermore, the code has been made compatible with means of automatic differentiation, enabling gradient-based optimization of nanophotonic systems~\cite{Asadova2025Dec}.

In this work, we extend this framework to acoustic multiple-scattering problems and introduce the Python-based software \textit{acoustotreams}. Using the \textit{T}-matrix in scalar spherical wave (SSW) or scalar cylindrical wave (SCW) bases, we model pressure-wave scattering by arbitrarily shaped scatterers, finite clusters, and periodic lattices with complex unit cells. Additionally, we implement rotations, translations, and basis transformations within these representations. To describe scattering in multilayered (stratified) systems, we employ the \textit{S}-matrix formalism in the scalar plane wave (SPW) basis, exploiting its stacking property via the Redheffer star product~\cite{Redheffer1959}.

Existing \textit{T}-matrix-based acoustic codes include MULTEL~\cite{Sainidou2005Mar} and AcSmuthi~\cite{AcSmuthi}, which are derived from their electromagnetic counterparts MULTEM~\cite{Stefanou1998Sep} and SMUTHI~\cite{Egel2021Oct}. MULTEL simulates acoustic responses of 2D lattices in the SSW basis and multilayer structures using the \textit{S}-matrix method, but is limited to spherical scatterers and single-scatterer unit cells. AcSmuthi supports both spherical and nonspherical scatterers, but only single ones, and can compute acoustic forces acting on them.

The paper is organized as follows. Section~\ref{sec:theory} presents the theoretical background, including the considered basis sets, the \textit{T}-matrix formulations for single and multiple scattering, the \textit{S}-matrix formalism for stratified media, and relevant operators. Section~\ref{sec:structure} describes the implementation and usage of the program. Finally, Section~\ref{sec:validation} provides validation results. Methods implemented in \textit{acoustotreams} have already been applied in studies on optimal multipole centers of acoustic scatterers~\cite{Ustimenko2025Apr}, bound states in the continuum in acoustic metasurfaces~\cite{Ustimenko2026Jan,Ustimenko2026Feb}, and retrieval of effective properties of ultrasonic metamaterials~\cite{Demeulenaere2025Dec}.

\begin{figure}[h!]
    \centering
    \includegraphics[width=\linewidth]{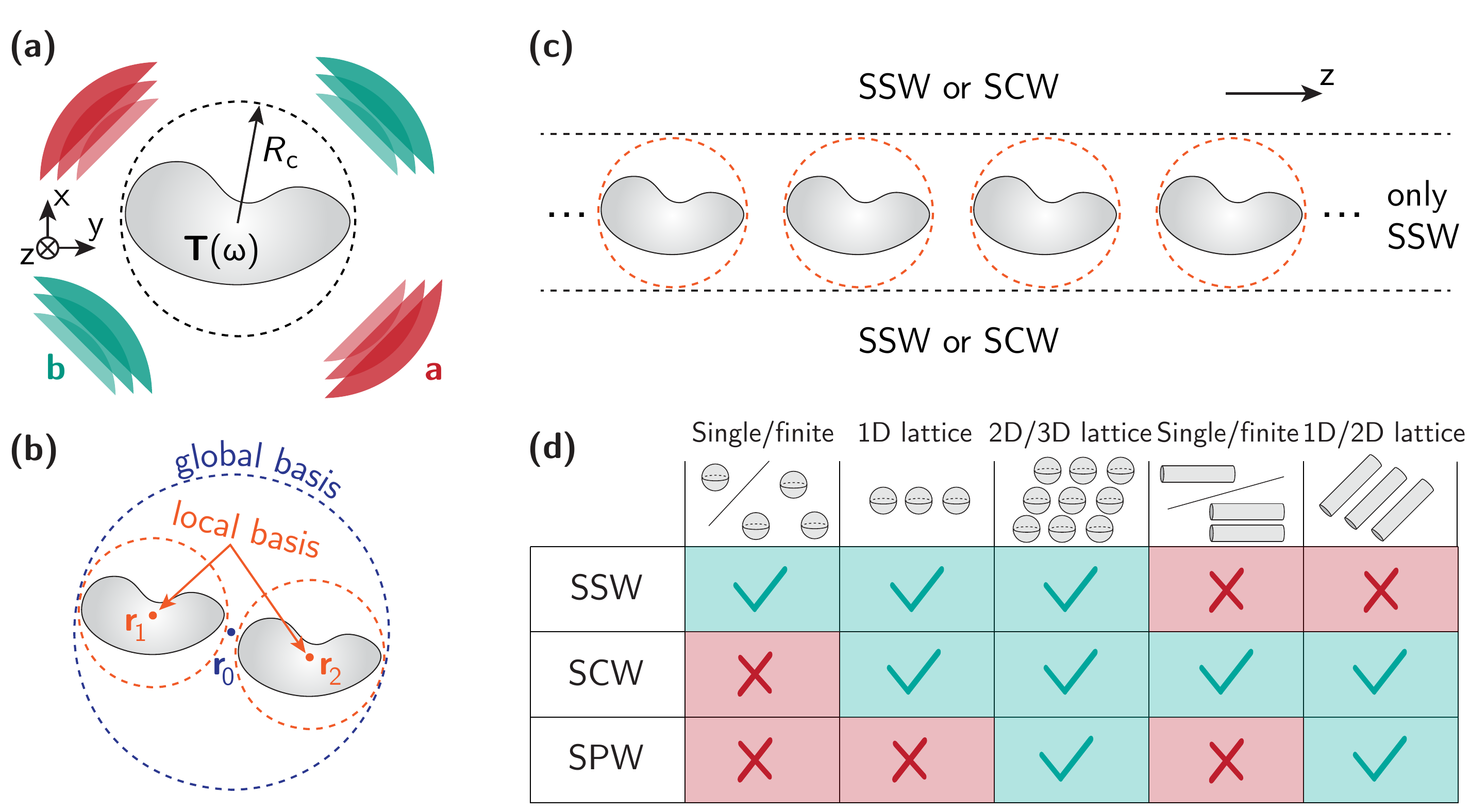}
    \caption{(a) A scatterer (shown in gray) transforms an incident wave (green) into a scattered wave (red). The scatterer can be finite or infinitely extended along the $z$-axis. The \textit{T}-matrix of the scatterer relates the incident ($\mathbf{b}$) and scattered ($\mathbf{a}$) field expansion coefficients in the SSW or SCW basis, respectively. Their validity domain is $|\mathbf{r}| > R_{\mathrm{c}}$, where $R_{\mathrm{c}}$ is the radius of the circumscribing sphere or infinite cylinder (dashed circle), respectively. (b) Orange and blue dashed lines depict the validity domains of the local (with multiple origins at $\mathbf{r}_1$ and $\mathbf{r}_2$) and global (with a single origin at $\mathbf{r}_0$) descriptions in a multiple-scattering problem. (c) The validity domains of the SSW and SCW bases for the description of scattering by a 1D lattice of finite 3D scatterers are shown by orange and black dashed lines. (d) Structures, for which the acoustic scattering can be simulated in \textit{acoustotreams} using the SSW, SCW, and SPW basis sets. Here, gray spheres and cylinders depict circumscribing spheres and cylinders, respectively, while the actual scatterers can have arbitrary shapes.}  
    \label{fig:main}
\end{figure}

\section{Theoretical formalism}\label{sec:theory}

This section introduces the governing equations of linear pressure acoustics, solved by \textit{acoustotreams}. This section defines the SSW, SCW, and SPW basis sets as solutions of the scalar Helmholtz equation in three dimensions, in spherical, cylindrical, and Cartesian coordinates, respectively. Using the SSW and SCW bases, acoustic scattering can be described via the \textit{T}-matrix [see Fig.~\ref{fig:main}(a)], which also enables the treatment of multiple-scattering problems.

For finite systems, both local and global scattering descriptions can be employed [see Fig.~\ref{fig:main}(b)]. For periodic systems (lattices), computational efficiency can be improved by selecting basis sets adapted to the structural symmetry. For example, Fig.~\ref{fig:main}(c) illustrates SSW and SCW-based descriptions of a 1D lattice.

Finally, the acoustic \textit{S}-matrix in the SPW basis is introduced to describe wave propagation and scattering in stratified media. Figure~\ref{fig:main}(d) summarizes the classes of structures that can be simulated using the basis sets implemented in \textit{acoustotreams}.

\subsection{The acoustic wave equation}

We consider a fluid medium characterized by an isotropic density $\rho$ and speed of sound $c$, in which only longitudinal (compressional) waves can propagate (in contrast to solids, which may also support shear waves). Acoustic waves are treated within first-order perturbation theory and are described by the conservation of linear momentum and mass~\cite{Williams1999,Bruus2011Dec}.

In the frequency domain, the governing equations are
\begin{align}
\label{main_eqs}
\begin{pmatrix}
\bm{\nabla}& 0 \\
0& \bm{\nabla} \cdot
\end{pmatrix}
\begin{pmatrix}
p\\
\mathbf{v}
\end{pmatrix} =
\mathrm i \omega
\begin{pmatrix}
0& \rho\\
\beta& 0
\end{pmatrix}
\begin{pmatrix}
p\\
\mathbf{v}
\end{pmatrix}\,,
\end{align}
where $\mathrm i = \sqrt{-1}$, $\beta = (\rho c^2)^{-1}$ is the compressibility and $\omega$ is the angular frequency. The fields $p(\omega; \mathbf{r})$ and $\mathbf{v}(\omega; \mathbf{r})$ denote the pressure and velocity at position $\mathbf{r} \in \mathbb{R}^3$. Measurable time-dependent fields are obtained as the real parts of the corresponding complex fields $\Re[p(\omega; \mathbf{r})\mathrm{e}^{-\mathrm{i} \omega t}]$ and $\Re[\mathbf{v}(\omega; \mathbf{r})\mathrm{e}^{-\mathrm{i} \omega t}]$. The time-dependent factor $\mathrm{e}^{-\mathrm{i} \omega t}$ is omitted hereafter for brevity.

From Eq.~\eqref{main_eqs}, the Helmholtz equations follow:
\begin{subequations}
   \begin{align}
   \label{eq_p_helmholtz}
    \Delta p + k^2 p &= 0\,, \\
    \label{eq_v_helmholtz}
    \bm{\nabla} ( \bm{\nabla} \cdot \mathbf{v}) + k^2 \mathbf{v} &= 0\,,
\end{align} 
\end{subequations}
with the angular wavenumber $k = \omega/c$.

\subsection{Scalar basis sets}\label{sec:basis}
We denote $\Psi_{\xi}(\omega; \mathbf{r})$ as a solution of Eq.~\eqref{eq_p_helmholtz}, where the subscript $\xi$ represents a general set of indices. From Eq.~\eqref{main_eqs}, the velocity field is related to the pressure field via $\mathbf{v} = (\mathrm{i}\omega\rho)^{-1}\bm{\nabla} p$. Accordingly, a corresponding solution of Eq.~\eqref{eq_v_helmholtz} is given by the longitudinal vector field $\mathbf{L}_{\xi}(\omega; \mathbf{r}) = k^{-1}\bm{\nabla}\Psi_{\xi}(\omega; \mathbf{r})$, where normalization by $k$ ensures dimensionless units.

Although transverse vector solutions can also be constructed from $\Psi_{\xi}(\omega; \mathbf{r})$, they are not required for velocity fields satisfying $\bm{\nabla} \times \mathbf{v} = \mathbf{0}$ and are, therefore, omitted here for brevity.

The explicit form of $\Psi_{\xi}(\omega; \mathbf{r})$ depends on the coordinate system in which the Laplacian $\Delta$ is expressed. In \textit{acoustotreams}, we implement analytical solutions in Cartesian, spherical, and cylindrical coordinates~\cite{Morse1953}. These solutions are explicitly discussed in the following.

\subsubsection{Plane waves}
In Cartesian coordinates $(x,y,z)$, a scalar plane wave is defined by the wave vector $\mathbf{k} = k_x\hat{\mathbf{x}} + k_y\hat{\mathbf{y}} + k_z\hat{\mathbf{z}}$,
\begin{subequations}
\begin{align}
\label{eq_spw_def}
    \Psi_{\mathbf{k}}(\omega; \mathbf{r}) = \mathrm{e}^{\mathrm{i}\mathbf{k}\mathbf{r}}\,.
\end{align}
The corresponding longitudinal vector plane wave is
\begin{align}
\label{eq_vpw_def}
    \mathbf{L}_{\mathbf{k}}(\omega; \mathbf{r}) = \mathrm{i} \hat{\mathbf{k}} \mathrm{e}^{\mathrm{i}\mathbf{k}\mathbf{r}}\,,
\end{align}
\end{subequations}
where $\widehat{\mathbf{k}} = \mathbf{k}/k$ and $k = \sqrt{\mathbf{k}\cdot\mathbf{k}}$.

In \textit{acoustotreams}, scalar plane waves can be specified either by all three components of $\mathbf{k}$ or, in the case of stratified media, by the in-plane component $\mathbf{k}_{\parallel}$. In the latter case, the remaining component is determined by the dispersion relation, with its sign set by the propagation direction (see Sec.~\ref{sec:smat}).

\subsubsection{Spherical waves}\label{sec:planewaves}
In spherical coordinates $(r,\theta,\varphi)$, scalar spherical waves (SSWs) are indexed by a degree $l \in \mathbb{N}_0$ and order $m \in \{-l, -l+1,\dots,l\}$:
\begin{subequations}
\begin{align}
\label{eq_ssw_def}
    \Psi^{(n)}_{l,m}(\omega; \mathbf{r}) = z^{(n)}_l(k r) Y_{l,m}(\theta, \varphi)\,.
\end{align}
For incident fields, spherical Bessel functions $z^{(1)}_l(kr) \equiv j_l(kr)$ are used due to their regularity at $kr=0$. For scattered fields, spherical Hankel functions of the first kind $z^{(3)}_l(kr) \equiv h^{(1)}_l(kr)$ are employed, as they satisfy the Sommerfeld radiation condition for $kr \to \infty$ but are singular at the origin. Such a singularity is not a problem because the scattered field is defined only outside the circumscribing sphere of the scatterer, and therefore not at the origin.
The scalar spherical harmonic is given by $Y_{l,m}(\theta,\varphi) = L_{l,m} P_l^m(\cos\theta)\mathrm{e}^{\mathrm{i}m\varphi}$, where $P_l^m$ is the associated Legendre polynomial.

Applying the gradient operator yields the longitudinal vector spherical wave (LVSW):
\begin{align}
\label{eq_vsw_def}
    \mathbf{L}_{l, m}^{(n)}(\omega; \mathbf{r}) = L_{l,m} \left[ z^{(n)\prime}_{l}(kr) P_{l}^m(\cos \theta) \hat{\mathbf{r}} + \frac{z^{(n)}_{l}(kr)}{kr} \left( \tau_{l,m}(\theta) \hat{\bm{\theta}} + \mathrm{i} \pi_{l,m}(\theta) \hat{\bm{\varphi}} \right) \right] \mathrm{e}^{\mathrm{i} m \varphi}\,,
\end{align}
\end{subequations}
where $\prime$ denotes the derivative of $z^{(n)}_{l}(kr)$ with respect to $(kr)$. The prefactor is
\begin{align}
    L_{l,m} = \sqrt{\frac{2l + 1}{4 \pi} \frac{(l - m)!}{(l + m)!}}\,,
\end{align}
and the angular functions are defined as
\begin{align}
    \tau_{l,m}(\theta) = \frac{\partial P_{l}^m(\cos \theta)}{\partial  \theta}\,,\quad
    \pi_{l,m}(\theta) = \frac{mP_{l}^m(\cos \theta)}{\sin \theta}\,.
\end{align}

\subsubsection{Cylindrical waves}
In cylindrical coordinates $(\rho,\varphi,z)$, scalar cylindrical waves (SCWs) are defined as
\begin{subequations}
\begin{align}
\label{eq_scw_def}
    \Psi^{(n)}_{k_z,m}(\omega; \mathbf{r}) = Z^{(n)}_m(k_{\rho}\rho) \mathrm{e}^{\mathrm{i} m \varphi + \mathrm{i} k_z z}\,,
\end{align}
where $k_z \in \mathbb{R}$ is the $z$ component of the wave vector, which determines the radial component as $k_{\rho} = \sqrt{k^2 - k_z^2}$, and $m \in \mathbb{Z}$ is the order. The functions $Z^{(n)}_m$ are Bessel ($n=1$) or Hankel ($n=3$) functions corresponding to regular and outgoing waves, respectively. 

The associated longitudinal vector cylindrical wave is
\begin{align}
\label{eq_vcw_def}
    \mathbf{L}_{k_z, m}^{(n)}(\omega;\mathbf{r}) &= \left[\frac{k_{\rho}}{k} Z^{(n)'}_{m}(k_{\rho}\rho) \hat{\bm{\rho}} + \mathrm{i}\frac{m k_{\rho}}{k}\frac{Z_{m}^{(n)}(k_{\rho}\rho)}{k_{\rho}\rho} \hat{\bm{\varphi}} + \frac{\mathrm{i} k_z}{k}Z_{m}^{(n)}(k_{\rho}\rho) \hat{\mathbf{z}} \right] \mathrm{e}^{\mathrm{i} m \varphi + \mathrm{i} k_z z}\,.
\end{align}
\end{subequations}

\subsection{Multipolar expansion}\label{sec:multipolar_expansion}
With the SSW and SCW bases being established, we consider acoustic scattering in a background fluid characterized by a density $\rho_{\mathrm{b}}$ and sound speed $c_{\mathrm{b}}$. These parameters are real-valued scalars. We also assume that the coordinate origin $\mathbf{r}=\mathbf{0}$ is placed at the scatterer’s center of mass.

Due to the linearity of Eqs.~\eqref{eq_p_helmholtz} and~\eqref{eq_v_helmholtz}, the total fields outside the circumscribing sphere or cylinder of the scatterer can be expressed as
    \begin{align}
        p_{\rm tot}(\omega; \mathbf{r}) = p_{\rm inc}(\omega; \mathbf{r}) + p_{\rm sca}(\omega; \mathbf{r})\,, \quad
         \mathbf{v}_{\rm tot}(\omega; \mathbf{r}) = \mathbf{v}_{\rm inc}(\omega; \mathbf{r}) + \mathbf{v}_{\rm sca}(\omega; \mathbf{r})\,.
    \end{align}
In the background medium, the incident and scattered acoustic pressure fields can be expanded into a series of regular and singular multipolar waves, respectively [see Fig.~\ref{fig:main}(a)], as
\begin{subequations}
\begin{align}
\label{eq_p_expansion}
\begin{aligned}
    p_{\rm inc}(\omega; \mathbf{r}) = \sum_{\xi} b_{\xi}\Psi^{(1)}_{\xi}(\omega; \mathbf{r}-\mathbf{r}_0)\,, \quad
    p_{\rm sca}(\omega; \mathbf{r}) = \sum_{\xi} a_{\xi}\Psi^{(3)}_{\xi}(\omega; \mathbf{r}-\mathbf{r}_0)\,,
\end{aligned}
\end{align}
and the velocity fields as
\begin{align}
\label{eq_v_expansion}
\begin{aligned}
    \mathbf{v}_{\rm inc}(\omega; \mathbf{r}) = -\frac{\mathrm{i}}{\rho_{\rm b}c_{\rm b}}\sum_{\xi} b_{\xi}\mathbf{L}^{(1)}_{\xi}(\omega; \mathbf{r}-\mathbf{r}_0)\,, \quad
    \mathbf{v}_{\rm sca}(\omega; \mathbf{r}) = -\frac{\mathrm{i}}{\rho_{\rm b}c_{\rm b}}\sum_{\xi} a_{\xi}\mathbf{L}^{(3)}_{\xi}(\omega; \mathbf{r}-\mathbf{r}_0)\,,
\end{aligned}
\end{align}
\end{subequations}
where $\xi = \{l,m\}$ for SSWs and $\xi = \{k_z,m\}$ for SCWs. 
In the general case, the expansion coefficients $b_{\xi}$ and $a_{\xi}$ depend, among other things, on the frequency $\omega$ and the origin of the expansion $\mathbf{r}_0$. Please note that both single-origin ($\mathbf{r}_0$) and multiple-origin $\{ \mathbf{r}_1, \mathbf{r}_2, ..., \mathbf{r}_N\}$ expansions are supported in \textit{acoustotreams}. The use of SSWs requires the object to be finite in all directions, while the SCWs are suitable for objects infinite along one axis (the $z$-axis in \textit{acoustotreams}). 

Following the discussion for electromagnetic scatterers~\cite{Auguie2016May}, we shall mention that the convergence of series in Eqs.~\eqref{eq_p_expansion} and~\eqref{eq_v_expansion} is guaranteed only outside the circumscribing sphere or cylinder of the scatterer in the SSW or SCW basis, respectively, i.e., for $|\mathbf{r} - \mathbf{r}_0| > R_{\mathrm{c}}$, where $R_{\mathrm{c}}$ is the radius, as shown in Fig.~\ref{fig:main}(a).

Moreover, the infinite series in Eqs.~\eqref{eq_p_expansion} and~\eqref{eq_v_expansion} must be truncated in practice, so that we only include the waves that mainly contribute to the scattering. Importantly, Ganesh, Hawkins, and Hiptmair have shown that the acoustic \textit{T}-matrix is a compact operator~\cite{Ganesh2012Oct}, ensuring convergence with increasing truncation parameters $l = l_{\rm max}$ for SSWs and $|m| = m_{\rm max}$ for SCWs. 

The choice of truncation parameters depends on the size parameter (the scatterer size relative to the wavelength in the background medium), the material parameters of the scatterer and background, and the presence of other scatterers or a substrate in proximity. For example, $l_{\rm max} = m_{\rm max} = 1$ is usually sufficient to describe acoustic scattering by single subwavelength objects~\cite{Wei2020Aug,Silva2014Nov,Toftul2019Oct,Long2020Jun,Kovacevich2021Oct,Smagin2023Oct}. For larger scatterers in a homogeneous medium, a rule of thumb could be $l_{\rm max}$ or $m_{\rm max} \approx \frac{1}{2}(k_{\rm b} a) + 1$, where $k$ is the wavenumber in the background medium, $a$ is the characteristic size of the scatterer (e.g., the diameter for a sphere) and the size parameter $(ka)$ is rounded up to the larger integer~\cite{Necada2021Jan}. It is also worth mentioning that sufficient truncation parameters $l_{\rm max}$ and $m_{\rm max}$ may depend on the expansion center $\mathbf{r}_0$. The optimal position, which minimizes the number of terms in the expansion and maximizes the accuracy, may differ from the center of mass~\cite{Ustimenko2025Apr}. In this work, for the sake of simplicity, we assume that the expansion center is always located at the center of mass and omit the argument $\mathbf{r}_0$ hereafter.

For SCWs, the set of included $k_z$ components must also be chosen. A single component may suffice for uniform structures, whereas multiple components ($n_{k_z}$) are required for systems with periodicity along $z$, corresponding to different diffraction orders.

\subsection{Acoustic \textit{T}-matrix of a single scatterer}
In the linear-response regime, the acoustic \textit{T}-matrix relates the expansion coefficients of the incident and scattered fields in Eqs.~\eqref{eq_p_expansion} and~\eqref{eq_v_expansion} as
\begin{align}
\label{eq_T_mat}
    \mathbf{a} = \mathbf{T}(\omega) \mathbf{b}\,,
\end{align}
where $\mathbf{b}$ and $\mathbf{a}$ are vectors of the corresponding coefficients~\cite{Waterman1969Jun,Waterman2009Jan}. Their lengths are $(l_{\rm max} + 1)^2$ for SSWs and $n_{k_z} (2 m_{\rm max} + 1)$ for SCWs. 

The acoustic \textit{T}-matrix can be computed semi-analytically for multilayered spheres~\cite{Sainidou2005Mar} or cylinders (see~\ref{sec_cylinder}). For arbitrarily shaped objects, it can be obtained using numerical methods such as the extended boundary condition method~\cite{Waterman1965Aug,Waterman1969Jun}, the null-field method~\cite{Martin2006Aug}, or the two-part algorithm~\cite{Ganesh2008Oct}, implemented in the package TMATROM3~\cite{Ganesh2022Mar}. Our FEM-based model for computing the acoustic \textit{T}-matrix in COMSOL Multiphysics is described in Ref.~\cite{tm_comsol}. To validate numerically computed \textit{T}-matrices, we impose the physical constraints, which are provided in~\ref{sec_comsol} for \textit{T}-matrices in the SSW basis. 

A sufficiently large \textit{T}-matrix fully characterizes the scattering at a given frequency. It provides access to quantities such as the number of states and density of states~\cite{Alevizaki2017Jun}, as well as rotation-averaged scattering and extinction cross sections (for SSWs) or widths (for SCWs) at a frequency $\omega$, obtained by adapting the corresponding electromagnetic expressions~\cite{Mishchenko1996May}:
\begin{align}
\label{eq_cross_avg}
\langle \sigma_{\rm sca} \rangle =\eta \, \text{tr} \left(\mathbf{T}^\dagger \mathbf{T}\right)\,, \quad\quad \langle \sigma_{\rm ext} \rangle = -\eta \, \Re \left(\text{tr} \, \mathbf{T}\right)\,,
\end{align}
where the prefactor is $\eta = 4\pi k^{-2}$ for SSWs and $\eta = 4 k^{-1} n_{k_z}^{-1}$ for SCWs, respectively, $\dagger$ denotes Hermitian conjugation and tr denotes matrix trace. 

For a given incident field $\mathbf{b}$, the \textit{T}-matrix yields the coefficients $\mathbf{a}$ [Eq.~\eqref{eq_T_mat}], which can be used to compute the scattered pressure field [Eq.~\eqref{eq_p_expansion}] and the scattered velocity field [Eq.~\eqref{eq_v_expansion}]. In addition, \ref{sec:radpattern} shows how to calculate the far-field amplitude of the scattered pressure field, as well as the scattered ($P_{\rm sca}$) and the extinction ($P_{\rm ext}$) power. Dividing $P_{\rm sca}$ and $P_{\rm ext}$ by the incident energy flux $I_0$ provides us with the scattering and extinction cross sections (resp. widths) at a frequency $\omega$~\cite{Martin2019Mar}:
  \begin{align}
\label{eq_sigma_sca}
    \sigma_{\rm sca}=\frac{\eta}{I_{0}} \mathbf{a}^\dagger  \mathbf{a}\,, \quad  
    \sigma_{\rm ext}=-\frac{\eta}{I_{0}} \Re \left[ \mathbf{b}^\dagger \mathbf{a}\right]\,,
\end{align}  
respectively, where the prefactor is $\eta = (2k^2)^{-1}$ for SSWs and $\eta = 2k^{-1}$ for SCWs, respectively. In \textit{acoustotreams}, $I_0$ for a plane pressure wave is defined as $I_{0} = 1/2$. The absorption cross section (resp. width) is then $\sigma_{\rm abs} = \sigma_{\rm ext} - \sigma_{\rm sca}$.

\subsection{Expansion of the regular waves in another basis}
To apply Eq.~\eqref{eq_T_mat}, it is useful to express the coefficients $\mathbf{b}$ explicitly for a given incident field. As the incident wave, we frequently consider a plane wave defined by Eq.~\eqref{eq_spw_def}. Its expansion in regular SSWs is~\cite{Sainidou2005Mar,Morse1953}
\begin{align}
\label{eq_pw_into_sw}
    \Psi_{\mathbf{k}}(\omega; \mathbf{r}) = 4 \pi \sum\limits_{l=0}^{+\infty} \sum_{m = -l}^{l} L_{l, m} \mathrm{i}^{l} P^m_{l}(\cos \theta_{\mathbf{k}}) \mathrm{e}^{-\mathrm{i}m \varphi_{\mathbf{k}}} \Psi^{(1)}_{l ,m}(\omega; \mathbf{r})\,,
\end{align}
and into regular SCWs~\cite{Morse1953}
\begin{align}
\label{eq_pw_into_cw}
    \Psi_{\mathbf{k}}(\omega; \mathbf{r}) = \sum\limits_{m=-\infty}^{+\infty} \mathrm{i}^{m}\mathrm{e}^{-\mathrm{i}m \varphi_{\mathbf{k}}} \Psi^{(1)}_{k_z, m}(\omega; \mathbf{r})\,.
\end{align}
Combining Eqs.~\eqref{eq_pw_into_sw} and~\eqref{eq_pw_into_cw} yields the expansion of a regular SCW in terms of regular SSWs:
\begin{align}
\label{eq_cw_into_sw}
    \Psi^{(1)}_{k_z, m}(\omega; \mathbf{r}) = 4 \pi \sum\limits_{l=|m|}^{+\infty}  L_{l, m} \mathrm{i}^{l-m} P^m_{l}(\cos \theta_{\mathbf{k}}) \Psi^{(1)}_{l, m}(\omega; \mathbf{r})\,.
\end{align}
In \textit{acoustotreams}, truncated versions of these expansions are implemented as transformation matrices between bases. Inverse transformations, which involve integrals over continuous spectra, are not implemented. However, their analytical expressions are provided in~\ref{sec_lattice_sums}.  

\subsection{\textit{T}-matrix method for finite clusters of scatterers}
\subsubsection{Local description}
The \textit{T}-matrix method is particularly powerful for multiple-scattering problems~\cite{Martin2006Aug,Li2024Aug,Peterson1975Jan,Kafesaki1999Nov}. We consider $N$ scatterers with their centers of mass located at $\mathbf{r}_i$, where $i = 1 ...N$. For accuracy, their circumscribing spheres or cylinders of radii $R_{\mathrm c}^{(i)}$ should not overlap [cf. the orange circles in Fig.~\ref{fig:main}(b)]

To find the scattered field coefficients, we note that each scatterer experiences both the external incident field and the fields scattered by all other scatterers. Writing the insonification as such a superposition leads to the linear system of equations
\begin{align}
\label{eq_mse}
    \mathbf{a}_i = \mathbf{T}_i(\omega) \left[ \mathbf{b}_i + \sum_{j \neq i} \bm{\mathcal{C}}^{(3)}_{ij}\mathbf{a}_j\right]\,,
\end{align}
where $\mathbf{b}_i$ contains the expansion coefficients of the external incident field with respect to the $i$th scatterer. The matrix $\bm{\mathcal{C}}^{(3)}_{ij} \equiv \bm{\mathcal{C}}^{(3)}\left(\omega;\mathbf{r}_i - \mathbf{r}_j\right)$ contains singular translation coefficients that expand regular incident waves at $\mathbf{r}_i$ in singular scattered waves at $\mathbf{r}_j$,
\begin{align}
\label{eq_transl_sing}
    \Psi^{(3)}_{\xi} (\omega; \mathbf{r}_i) = \sum_{\xi'} \mathcal{C}_{\xi' \xi}^{(3)}(\omega;\mathbf{r}_i-\mathbf{r}_j)\Psi^{(1)}_{\xi'} (\omega; \mathbf{r}_j)\,,
\end{align}
where $\mathcal{C}_{\xi' \xi}^{(3)}(\omega;\mathbf{r})$ are provided in~\ref{translation_coeffs}. Combining vectors $\mathbf{a}_i$ and $\mathbf{b}_i$ into column vectors $\mathbf{a} = \left[\mathbf{a}_1, \dots,  \mathbf{a}_N\right]$ and 
$\mathbf{b}^{\rm local} = \left[\mathbf{b}_1, \dots,  \mathbf{b}_N\right]$ yields
\begin{align}
\label{eq_mse_local}
    \mathbf{a} = \mathbf{T}^{\rm diag}(\omega)\left[\mathbf{b}^{\rm local} +  \bm{\mathcal{C}}^{(3)}\mathbf{a}\right]\,,
\end{align}
where the following block-diagonal matrices have been introduced 
\begin{align}
\label{eq_Tdiag}
    \mathbf{T}^{\rm diag} = 
    \begin{pmatrix}
        \mathbf{T}_1& \bm{0}& \dots & \bm{0} \\
        \bm{0}& \mathbf{T}_2&  \dots & \bm{0} \\
        \vdots& \vdots& \ddots& \vdots \\
        \bm{0}& \bm{0}& \dots & \mathbf{T}_N
    \end{pmatrix}\,, \quad
    \bm{\mathcal{C}}^{(3)} =
    \begin{pmatrix}
        \bm{0}& \bm{\mathcal{C}}^{(3)}_{12}& \dots & \bm{\mathcal{C}}^{(3)}_{1N}\\
        \bm{\mathcal{C}}^{(3)}_{21}& \bm{0}&  \dots & \bm{\mathcal{C}}^{(3)}_{2N}\\
        \vdots& \vdots& \ddots& \vdots \\
        \bm{\mathcal{C}}^{(3)}_{N1}& \bm{\mathcal{C}}^{(3)}_{N2}& \dots &  \bm{0}
    \end{pmatrix}\,.
\end{align}
Note that the diagonal entries of $\bm{\mathcal{C}}^{(3)}$ vanish to exclude self-interaction, which is taken into account by the \textit{T}-matrices.  

Solving Eq.~\eqref{eq_mse_local} for $\mathbf{a}$ gives
\begin{subequations}
    \begin{align}
    \label{eq_pTa_local}
    \mathbf{a}^{\rm local} = \mathbf{T}^{\rm local}(\omega)\mathbf{b}^{\rm local}\,,
\end{align}
where the matrix
\begin{align}
\label{eq_T_local}
    \mathbf{T}^{\rm local}(\omega) = \left[ \mathbf{I} - \mathbf{T}^{\rm diag}(\omega)\bm{\mathcal{C}}^{(3)} \right]^{-1}\mathbf{T}^{\rm diag}(\omega)\,,
\end{align}
\end{subequations}
is the \textit{T}-matrix in a so-called local basis, which considers the mutual interaction between the objects self-consistently [see Fig.~\ref{fig:main}(b)], and $\mathbf{I}$ is an identity matrix of the corresponding size.
The coefficients in Eq.~\eqref{eq_pTa_local} determine the total pressure field as
\begin{align}
    p_{\rm tot}(\omega; \mathbf{r}) = p_{\rm inc}(\omega; \mathbf{r}) + \sum_{i = 1}^N \sum_{\xi} a^{\rm local}_{i,\xi}\Psi^{(3)}_{\xi}(\omega; \mathbf{r}-\mathbf{r}_i)\,,
\end{align}
where $\mathbf{a}^{\rm local}_i$ refers to the part of the vector $\mathbf{a}_{\rm local}$ corresponding to the $i$th scatterer. Convergence of this expansion to the actual field is guaranteed if $|\mathbf{r}-\mathbf{r}_i| > R_{\mathrm c}^{(i)}$ for any $i$ [cf. the orange circles in Fig.~\ref{fig:main}(b)].

\subsubsection{Global description}
Alternatively, the cluster can be described using a single expansion center $\mathbf{r}_0$ as illustrated in Fig.~\ref{fig:main}(b)~\cite{Suryadharma2017Jul}. In this case, the circumscribing sphere encapsulates the entire cluster [cf. the blue circle in Fig.~\ref{fig:main}(b)]. The scattered-field coefficients are transformed as
\begin{align}
\label{eq_loc_to_glob_sca}
\begin{aligned}
    \mathbf{a}^{\rm global} =  \sum_{i=1}^N \boldsymbol{\mathcal{C}}^{(1)}(\omega; \mathbf{r}_0 -\mathbf{r}_i)\mathbf{a}^{\rm local}_i \equiv \left[ \bm{\mathcal{C}}^{(1)}_{01} ... \bm{\mathcal{C}}^{(1)}_{0N}\right]\mathbf{a}^{\rm local}\,.
\end{aligned}
\end{align}
The matrix $\bm{\mathcal{C}}^{(1)}_{0i} \equiv \bm{\mathcal{C}}^{(1)}\left(\omega;\mathbf{r}_0 - \mathbf{r}_i\right)$ expands regular incident waves in regular incident waves ($n =1$) or singular scattered waves in singular scattered waves ($n =3$), using regular translation coefficients $\mathcal{C}_{\xi' \xi}^{(1)}(\omega;\mathbf{r}_0-\mathbf{r}_i)$ defined in~\ref{translation_coeffs},
\begin{align}
\label{eq_transl_reg}
    \Psi^{(n)}_{\xi} (\omega; \mathbf{r}_0) = \sum_{\xi'} \mathcal{C}_{\xi' \xi}^{(1)}(\omega;\mathbf{r}_0-\mathbf{r}_i)\Psi^{(n)}_{\xi'} (\omega; \mathbf{r}_i)\,.
\end{align}

Similarly, the incident-field coefficients are transformed as
\begin{align}
\label{eq_loc_to_glob_inc}
\begin{aligned}
\mathbf{b}_i^{\rm local} = \bm{\mathcal{C}}^{(1)}_{i0}\mathbf{b}^{\rm global}\, ,
\end{aligned}
\end{align}
where $\bm{\mathcal{C}}^{(1)}_{i0}  \equiv \bm{\mathcal{C}}^{(1)}\left(\omega;\mathbf{r}_i - \mathbf{r}_0\right)$. Combining Eqs.~\eqref{eq_loc_to_glob_sca} and~\eqref{eq_loc_to_glob_inc} yields
\begin{subequations}
    \begin{align}
    \label{eq_pTa_global}
    \mathbf{a}^{\rm global} = \mathbf{T}^{\rm global}(\omega)\mathbf{b}^{\rm global}\, ,
\end{align}
with a global \textit{T}-matrix given by 
\begin{align}
\label{eq_T_global}
    \mathbf{T}^{\rm global}(\omega) = \left[ \bm{\mathcal{C}}^{(1)}_{01} ... \bm{\mathcal{C}}^{(1)}_{0N}\right] \mathbf{T}^{\rm local}(\omega)\left[ \bm{\mathcal{C}}^{(1)}_{10} ... \bm{\mathcal{C}}^{(1)}_{N0}\right]\, .
\end{align}
\end{subequations}
Equation~\eqref{eq_pTa_global} is essentially the same as Eq.~\eqref{eq_T_mat}. Therefore, \textit{acoustotreams} treats global \textit{T}-matrices as local ones with a single expansion center (position) given. Using the global \textit{T}-matrix in Eq.~\eqref{eq_T_global}, we can calculate the rotation-averaged cross sections (resp. widths) in Eq.~\eqref{eq_cross_avg} for the cluster.

\subsubsection{Cross sections and cross widths}
To compute the scattering and extinction cross sections (resp. widths) in a global basis, we can rewrite Eq.~\eqref{eq_sigma_sca} with the global quantities
  \begin{align}
\label{eq_sigma_sca_global}
    \sigma_{\rm sca}^{\rm global}=\frac{\eta}{I_{0}} \left[\mathbf{a}^{\rm global}\right]^\dagger  \mathbf{a}^{\rm global}\,, \quad 
    \sigma_{\rm ext}^{\rm global}=-\frac{\eta}{I_{0}} \Re \left( \left[\mathbf{b}^{\rm global}\right]^\dagger \mathbf{a}^{\rm global}\right)\,.
\end{align}  
If we plug Eqs.~\eqref{eq_loc_to_glob_sca} and~\eqref{eq_loc_to_glob_inc} into Eq.~\eqref{eq_sigma_sca_global}, we obtain the cross sections (resp. widths) in a local basis
  \begin{align}
\label{eq_sigma_sca_local}
    \sigma_{\rm sca}^{\rm local} = \frac{\eta}{I_{0}} \sum_{i,j=1}^N\left[\mathbf{a}^{\rm local}_i\right]^\dagger  \bm{\mathcal{C}}_{ij}^{(1)} \mathbf{a}^{\rm local}_j\, , \quad  
    \sigma_{\rm ext}^{\rm local} =-\frac{\eta}{I_{0}} \Re \left( \left[\mathbf{b}^{\rm local}\right]^\dagger \mathbf{a}^{\rm local}\right)\, ,
\end{align}  
where $\bm{\mathcal{C}}_{ij}^{(1)} \equiv \bm{\mathcal{C}}^{(1)}(\omega; \mathbf{r}_i - \mathbf{r}_j)$ and the following properties of the regular translation coefficients have been used~\cite{Kim2004}:
\begin{subequations}
    \begin{align}
        \left[ \bm{\mathcal{C}}^{(1)}(\omega; \mathbf{r}_0 - \mathbf{r}_i)\right]^\dagger = \bm{\mathcal{C}}^{(1)}(\omega; \mathbf{r}_i - \mathbf{r}_0)\, ,
    \end{align}
    and
    \begin{align}
        \bm{\mathcal{C}}^{(1)}(\omega; \mathbf{r}_i - \mathbf{r}_j) = \bm{\mathcal{C}}^{(1)}(\omega; \mathbf{r}_i - \mathbf{r}_0) \bm{\mathcal{C}}^{(1)}(\omega; \mathbf{r}_0 - \mathbf{r}_j)\, . 
    \end{align}
\end{subequations}
Please note that only the expressions in Eq.~\eqref{eq_sigma_sca_local} are implemented in \textit{acoustotreams}, since $\bm{\mathcal{C}}_{ii}^{(1)} \equiv \mathbf{I}$ for $N = 1$. Moreover, the fields and cross sections in the local and global bases are equal if the global basis contains a sufficiently large number of multipolar waves to ensure convergence.

\subsection{\textit{T}-matrix method for periodic arrangements of scatterers}
In this subsection, we consider periodic arrangements, or lattices, of scatterers. A distinctive feature of \textit{acoustotreams} is (1) its support for complex unit cells, i.e., unit cells containing multiple scatterers located at positions $\mathbf{r}_i$, and (2) its use of the Ewald summation technique to efficiently evaluate lattice sums, essential for describing scattering in periodic systems within the \textit{T}-matrix and other multipole-based frameworks. These computations are carried out using functions from the subpackage \texttt{treams.lattice}.

\subsubsection{Effective \textit{T}-matrix and lattice sums}
Let us consider an \textit{s}-dimensional lattice $\Lambda_s$ defined by vectors $\mathbf{R} = \sum_{q=1}^{s} n_q \hat{\mathbf{e}}_q$, where $n_q \in \mathbb{Z}$ and $\widehat{\mathbf{e}}_q$ are the basis vectors. In \textit{acoustotreams}, a 2D lattice of SSWs is, by default, placed in the $xy$-plane, a 1D lattice of SSWs is aligned along the $z$-axis, and a 1D lattice of SCWs is aligned along the $x$-axis. We assume that each unit cell contains $N$ scatterers located at positions $\left(\mathbf{r}_i + \mathbf{R}\right)$ for the cell indexed by $\mathbf{R}$, where $i = 1,\dots,N$.

Exploiting the translational symmetry of the lattice, we invoke Bloch’s theorem and restrict the analysis to a single unit cell, called the reference unit cell, which we take at $\mathbf{R} = \mathbf{0}$. The scattered-field expansion coefficients for this unit cell follow from Eq.~\eqref{eq_mse} with an additional summation over all lattice sites:
\begin{align}
\label{eq_mse_pbc}
    \mathbf{a}_i = \mathbf{T}_i(\omega) \left[ \mathbf{b}_i + \sum_{j \neq i} \bm{\Sigma}_{ij}(\omega, \mathbf{k}_{\parallel})\mathbf{a}_j\right]\,.
\end{align}
The lattice sums of singular translation coefficients are defined as
\begin{align}
\label{eq_lattice_sums_def}
   \bm{\Sigma}_{ij}(\omega, \mathbf{k}_{\parallel}) = 
   \begin{cases}
          \sum\limits_{\mathbf{R} \in \Lambda_s} \bm{\mathcal{C}}^{(3)}\left(\omega; \mathbf{r} - \mathbf{r}_j - \mathbf{R}\right) \mathrm{e}^{\mathrm{i} \mathbf{k}_{\parallel}\mathbf{R}}\,, &\mathbf{r}\neq \mathbf{r}_j\,, \\
          \sum\limits_{\mathbf{R} \in \Lambda_s, \mathbf{R}\neq\mathbf{0}} \bm{\mathcal{C}}^{(3)}\left(\omega; - \mathbf{R}\right) \mathrm{e}^{\mathrm{i} \mathbf{k}_{\parallel}\mathbf{R}}\,, &\mathbf{r} = \mathbf{r}_j\,,
   \end{cases}
\end{align}
where the factor $\mathrm{e}^{\mathrm{i} \mathbf{k}_{\parallel}\mathbf{R}}$ accounts for the phase shift between unit cells at $\mathbf{0}$ and $\mathbf{R}$, as dictated by Bloch’s theorem.

We then assemble the matrix $\bm{\Sigma}$ from the blocks $\bm{\Sigma}_{ij}$. Its structure is similar to that of $\bm{\mathcal{C}}^{(3)}$ in Eq.~\eqref{eq_Tdiag}, except that its diagonal entries are nonzero, reflecting interactions between a scatterer at $\mathbf{r}_i$ and scatterers at $\left(\mathbf{r}_i + \mathbf{R}\right)$:
\begin{align}
\bm{\Sigma} =
    \begin{pmatrix}
       \bm{\Sigma}_{11}&\bm{\Sigma}_{12}& \dots & \bm{\Sigma}_{1N}\\
       \bm{\Sigma}_{21}& \bm{\Sigma}_{22}&  \dots & \bm{\Sigma}_{2N}\\
        \vdots& \vdots& \ddots& \vdots \\
        \bm{\Sigma}_{N1}& \bm{\Sigma}_{N2}& \dots &  \bm{\Sigma}_{NN}
    \end{pmatrix}\,.
\end{align}
Analogously to Eqs.~\eqref{eq_pTa_local} and~\eqref{eq_T_local}, the solution of Eq.~\eqref{eq_mse_pbc} can be written as
\begin{subequations}
    \begin{align}
\label{eq_pTa_eff}
    \mathbf{a}^{\rm eff} = \mathbf{T}^{\rm eff}(\omega, \mathbf{k}_{\parallel})\mathbf{b}\,,
\end{align}
where the effective \textit{T}-matrix is given by
\begin{align}
\label{eq_Teff}
    \mathbf{T}^{\rm eff}(\omega, \mathbf{k}_{\parallel}) = \left[ \mathbf{I} - \mathbf{T}^{\rm diag}(\omega)\bm{\Sigma}(\omega, \mathbf{k}_{\parallel}) \right]^{-1}\mathbf{T}^{\rm diag}(\omega)\,.
\end{align}
\end{subequations}
The matrix $\mathbf{T}^{\rm eff}$ has the same dimension as $\mathbf{T}^{\rm local}$ in Eq.~\eqref{eq_T_local}, but additionally incorporates interactions between all unit cells. Importantly, it depends not only on the frequency $\omega$ but also on the wave vector $\mathbf{k}_{\parallel}$ through the lattice sums. This dependence can be removed by integrating over all $\mathbf{k}_{\parallel}$~\cite{Zerulla2023Feb}. Please note that the basis for all quantities in Eq.~\eqref{eq_pTa_eff} can only be local.

\subsubsection{Condition of lattice eigenmodes}
Equation~\eqref{eq_pTa_eff} can be used not only to compute the response of the lattice to an external excitation but also to determine its eigenmodes in the absence of external stimuli~\cite{Chaplain2025Jan}. Setting $\mathbf{b} = \mathbf{0}$ yields
\begin{align}
\label{eq_Teff_eigenmodes}
    \left[ \mathbf{T}^{\rm eff}(\omega, \mathbf{k}_{\parallel}) \right]^{-1} \mathbf{a}^{\rm eff} = \mathbf{0}\,.
\end{align}
Thus, an eigenmode $\mathbf{a}^{\rm eff}$ corresponds to an eigenvector of the inverse effective \textit{T}-matrix with a vanishing (or sufficiently small) eigenvalue~\cite{Bulgakov2014Nov}, or equivalently to a right singular vector associated with a zero singular value~\cite{Necada2021Jan}. The pairs $(\omega,\mathbf{k}_{\parallel})$ for which Eq.~\eqref{eq_Teff_eigenmodes} admits nontrivial solutions define the dispersion relation of the periodic system. If the imaginary part of the eigenfrequency contains no radiative contribution, the mode becomes trapped, and it is then referred to as a bound state in the continuum, which can be efficiently identified via the singular value decomposition of the effective \textit{T}-matrix~\cite{Ustimenko2026Feb}.

\subsubsection{Scattered field of a lattice}
Using Eq.~\eqref{eq_pTa_eff}, the lattice sums~\eqref{eq_lattice_sums_def}, and the Bloch theorem, the total scattered field from a lattice can be written as
\begin{align}
\label{eq_psca_array}
    p_{\mathrm{sca}}(\omega; \mathbf{r}) = \sum_{i=1}^{N} \sum\limits_{\xi} a^{\rm eff}_{i,\xi}\sum_{\mathbf{R} \in \Lambda_s} \Psi^{(3)}_{\xi}(\omega; \mathbf{r} - \mathbf{r}_i  - \mathbf{R}) \mathrm{e}^{\mathrm{i} \mathbf{k}_{\parallel}\mathbf{R}}\,.
\end{align}
To evaluate the lattice sum of singular waves, we use its translation properties analogous to those employed in electromagnetic solvers~\cite{Beutel2024Apr,Necada2021Jan}. Specifically, for each $i$, the sum in Eq.~\eqref{eq_psca_array} can be re-expanded in terms of regular waves using Eq.~\eqref{eq_lattice_sums_def}:
\begin{align}
\label{eq_lattice_reexpansion}
\begin{aligned}
    \sum_{\mathbf{R} \in \Lambda_s} \Psi^{(3)}_{\xi}(\omega; \mathbf{r} - \mathbf{r}_i  - \mathbf{R}) \mathrm{e}^{\mathrm{i} \mathbf{k}_{\parallel}\mathbf{R}} = \sum\limits_{\xi'} \underbrace{\sum_{\mathbf{R} \in \Lambda_s} \mathcal{C}^{(3)}_{\xi \xi'}(\omega; \mathbf{r} - \mathbf{r}_i  - \mathbf{R})\mathrm{e}^{\mathrm{i} \mathbf{k}_{\parallel}\mathbf{R}}}_{\Sigma_{i,\xi,\xi'}(\omega, \mathbf{k}_{\parallel};\mathbf{r})} \times \Psi^{(1)}_{\xi'}(\omega; \mathbf{0})\,,
\end{aligned}
\end{align}
for $\mathbf{r} \neq \mathbf{r}_i$. Since $\Psi^{(1)}_{\xi'}(\omega; \mathbf{0})$ vanishes for most indices $\xi'$, the expression simplifies considerably. The only nonzero contributions are $\Psi_{0,0}^{(1)}(\omega;\mathbf{0}) = \frac{1}{\sqrt{4\pi}}$ for SSWs and $\Psi_{k_z,0}^{(1)}(\omega;\mathbf{0}) = 1$ for SCWs. Consequently, Eq.~\eqref{eq_psca_array} reduces to
\begin{align}
\label{eq_psca_lattice}
   p_{\mathrm{sca}}(\omega; \mathbf{r}) =
   \begin{cases}
       \dfrac{1}{\sqrt{4\pi}}\sum\limits_{i=1}^{N} \sum\limits_{l,m} a^{\rm eff}_{i,l,m}\Sigma_{i,l,m,0,0}(\omega, \mathbf{k}_{\parallel};\mathbf{r})\,, & \text{for SSWs}\,, \\
       \sum\limits_{i=1}^{N} \sum\limits_{k_z,m} a^{\rm eff}_{i,k_z,m}\Sigma_{i,k_z,m,k_z,0}(\omega, \mathbf{k}_{\parallel};\mathbf{r})\,, & \text{for SCWs}\,.
   \end{cases}
\end{align}
The velocity field can be computed analogously, noting that only $\mathbf{L}^{(1)}_{1,m}(\omega,\mathbf{0}) \neq 0$ in the SSW basis and $\mathbf{L}^{(1)}_{k_z,m}(\omega,\mathbf{0}) \neq 0$ with $|m|\leq1$ in the SCW basis (see Sec.~\ref{sec_lattice_field}).
 
\subsubsection{Expansion of the lattice sums in another basis}
We now present the expressions that convert lattice sums between different bases, offering an alternative route to evaluate Eq.~\eqref{eq_psca_array}. These expressions are derived using Fourier transforms of the SSWs and SCWs combined with the Poisson summation formula, which converts sums over the direct lattice $\Lambda_s$ into sums over the reciprocal lattice $\Lambda_s^{\ast}$ (see Sec.~\ref{sec_lattice_sums}).

Thus, a lattice sum of singular SSWs over a 1D lattice with a lattice constant of $L$, $\Lambda_1 = \left\{ n L \hat{\mathbf{z}} \big{|} n \in \mathbb{Z}\right\}$, can be converted into the sum of singular SCWs as
\begin{align}
\label{eq_sw_into_cw_lattice}
\begin{aligned}
    \sum_{\mathbf{R} \in \Lambda_1} \Psi^{(3)}_{l, m}(\omega; \mathbf{r} - \mathbf{R})\mathrm{e}^{\mathrm{i} \mathbf{k}_{\parallel} \mathbf{R}} = \frac{\pi L_{l, m}}{L k \mathrm{i}^{l-m}} \sum_{\mathbf{G} \in \Lambda^{\ast}_1} P^m_{l}(\cos \theta_{\mathbf{k}}) \Psi^{(3)}_{k_{\parallel} + G,m}(\omega;\mathbf{r})\,,
\end{aligned}
\end{align}
where, for a 1D lattice, $\mathbf{R} = nL\hat{\mathbf{z}} \in \Lambda_1$ goes over the direct lattice nodes, $G\hat{\mathbf{z}} = n\frac{2 \pi}{L}\hat{\mathbf{z}} \in \Lambda_1^{\ast}$ is the reciprocal lattice vector, $\mathbf{k}_{\parallel} = k_{\parallel} \hat{\mathbf{z}}$, and $\cos \theta_{\mathbf{k}} = \frac{k_{\parallel} + G}{k}$.

Furthermore, a sum of singular SSWs over a 2D lattice in the $xy$-plane can be converted into the sum of SPWs, which is used in particular to create the \textit{S}-matrix of a lattice from its effective \textit{T}-matrix in the SSW basis, using the following equation
\begin{align}
\label{eq_sw_into_pw_lattice}
\begin{aligned}
    \sum_{\mathbf{R} \in \Lambda_2} \Psi^{(3)}_{l, m}(\omega; \mathbf{r} - \mathbf{R})\mathrm{e}^{\mathrm{i} \mathbf{k}_{\parallel} \mathbf{R}} =\frac{2 \pi L_{l, m}}{A k^2 \mathrm{i}^{l}} \sum_{\mathbf{G} \in \Lambda^{\ast}_2} P^m_{l}(\cos \theta_{\mathbf{k}}) \frac{\mathrm{e}^{\mathrm{i}m \varphi_{\mathbf{k}}}}{\sqrt{1 - \frac{(\mathbf{k}_{\parallel} + \mathbf{G})^2}{k^2}}} \Psi_{\mathbf{k}_{\parallel} + \mathbf{G},d}(\omega;\mathbf{r})\,,
\end{aligned}
\end{align}
where, for a 2D lattice, $A$ is the area of a unit cell, $d = \pm1$ for waves propagating in the $z \gtrless 0$ direction, and $\cos \theta_{\mathbf{k}} =  \frac{d\sqrt{k^2 - (\mathbf{k}_{\parallel} + \mathbf{G})^2}}{k}$.

Finally, a sum of singular SCWs over a 1D lattice along the $x$-axis can be converted into the sum of SPWs as follows
\begin{align}
\label{eq_cw_into_pw_lattice}
\begin{aligned}
    \sum_{\mathbf{R} \in \Lambda_1} \Psi^{(3)}_{k_z, m}(\omega; \mathbf{r} - \mathbf{R})\mathrm{e}^{\mathrm{i} \mathbf{k}_{\parallel} \mathbf{R}} = \frac{2}{L k \mathrm{i}^{m}} \sum_{\mathbf{G} \in \Lambda^{\ast}_1}  \frac{ \mathrm{e}^{\mathrm{i}m \varphi_{\mathbf{k}}}}{\sqrt{1 - \frac{k_z^2 +(k_{\parallel} + G)^2}{k^2}}}\Psi_{\mathbf{k}}(\omega;\mathbf{r})\,,
\end{aligned}
\end{align}
where $\mathbf{k} = (k_{\parallel} + G) \hat{\mathbf{x}} + d\sqrt{k^2 - k_z^2 - (k_{\parallel} + G)^2} \hat{\mathbf{y}} + k_z \hat{\mathbf{z}}$. 

These transformations can remarkably accelerate computations by allowing one to choose a geometry-adapted basis. For example, one may convert the lattice sum for a 1D lattice evaluated in the SSW basis into the SCW basis using Eq.~\eqref{eq_sw_into_cw_lattice}, retaining only a few diffraction orders, and thus reducing the computational cost (see Sec.~\ref{sec:validation_lattice}). The price to pay is a reduced convergence domain: SSW expansions converge outside circumscribing spheres of individual scatterers, whereas SCW expansions converge only outside the circumscribing cylinder of the entire lattice and are, therefore, unsuitable for evaluating fields between scatterers [see Fig.~\ref{fig:main}(c)]. The use of different basis sets is illustrated in Fig.~\ref{fig:main}(d) and discussed in more detail in Ref.~\cite{Beutel2024Apr}.

\subsection{\textit{S}-matrices for wave propagation in stratified media}\label{sec:smat}
The \textit{T}-matrix relates the incident and scattered field coefficients in the SSW or SCW bases. To describe wave propagation in stratified media (infinite in the $xy$-plane), it is convenient to introduce the scattering matrix (\textit{S}-matrix), which connects incoming and outgoing plane-wave coefficients along the $z$-direction in the SPW basis. To this end, we write the wave vector of a SPW as $\mathbf{k} = \mathbf{k}_{\parallel} \pm k_z \hat{\mathbf{z}}$, where $\mathbf{k}_{\parallel} = k_x \hat{\mathbf{x}} + k_y \hat{\mathbf{y}}$ is the in-plane component, which is preserved during propagation, and $k_z = \sqrt{k^2 - k_{\parallel}^2}$ is the out-of-plane component (real for propagating waves and complex for evanescent waves). The ``$+$'' and ``$-$'' signs correspond to upward ($z>0$) and downward ($z<0$) propagation, respectively. The incoming and outgoing pressure fields are expanded in SPWs as
\begin{align}
\label{eq_p_to_a_pws}
\begin{aligned}
    p_{\rm in}(\omega; \mathbf{r}) &= 
    \begin{cases}
        \sum\limits_{\mathbf{k}_{\parallel}}  b_{\mathbf{k}_{\parallel},\uparrow} \Psi_{\mathbf{k}_{\parallel},\uparrow}(\omega; \mathbf{r})\,,& z <0\,, \\
        \sum\limits_{\mathbf{k}_{\parallel}}  b_{\mathbf{k}_{\parallel},\downarrow} \Psi_{\mathbf{k}_{\parallel},\downarrow}(\omega; \mathbf{r})\,,& z > 0\,,
    \end{cases}
     \\
     p_{\rm out}(\omega; \mathbf{r}) &= 
    \begin{cases}
        \sum\limits_{\mathbf{k}_{\parallel}}  a_{\mathbf{k}_{\parallel},\downarrow} \Psi_{\mathbf{k}_{\parallel},\downarrow}(\omega; \mathbf{r})\,,& z <0\,, \\
        \sum\limits_{\mathbf{k}_{\parallel}}  a_{\mathbf{k}_{\parallel},\uparrow} \Psi_{\mathbf{k}_{\parallel},\uparrow}(\omega; \mathbf{r})\,,& z > 0\,.
    \end{cases}
\end{aligned}
\end{align}
The \textit{S}-matrix relates the expansion coefficient vectors at $z=0$ as
\begin{align}
\label{eq_pSa}
    \begin{pmatrix}
        \mathbf{a}_{\uparrow} \\
        \mathbf{a}_{\downarrow}
    \end{pmatrix}
    =
    \underbrace{\begin{pmatrix}
        \mathbf{S}_{\uparrow\uparrow}& \mathbf{S}_{\uparrow\downarrow} \\
        \mathbf{S}_{\downarrow\uparrow}& \mathbf{S}_{\downarrow\downarrow}
    \end{pmatrix}}_{\mathbf{S}}
       \begin{pmatrix}
        \mathbf{b}_{\uparrow} \\
        \mathbf{b}_{\downarrow}
    \end{pmatrix}\,.
\end{align}
The diagonal blocks describe transmission, while the off-diagonal blocks correspond to reflection. For metasurfaces, the following aspects should be taken into account:
\begin{enumerate}
    \item The formulation is formally valid for $|z| > z_{\rm c}/2$, where $z_{\rm c}$ is the metasurface thickness.
    \item The \textit{S}-matrix is computed from the effective \textit{T}-matrix in the SSW or SCW basis. For the case of the SCW basis, \textit{acousotreams} automatically permutes the coordinate axes, so that the propagation axis is in the $z$ direction.
\end{enumerate}

\textit{S}-matrices can be stacked along the $z$-axis using the Redheffer star product~\cite{Redheffer1959}. For a periodic stack of \textit{S}-matrices in the $z$ direction, the band structure can be computed by imposing Bloch-periodic boundary conditions on the \textit{S}-matrix of a unit cell and solving the corresponding eigenvalue problem (see Sec.~2.4 in Ref.~\cite{Beutel2024thesis}).

\subsection{Other operators in \textit{acoustotreams}}\label{sec:other}

This subsection describes the remaining operators. First, arbitrary rotations of SSWs $\mathcal{R}(\alpha,\beta,\gamma)$ defined by Euler angles $\alpha,\beta,$ and $\gamma$ in the $zyz$ convention, as well as rotations of SCWs and SPWs about the $z$-axis. For SSWs, only the angular part $Y_{l,m}(\theta,\varphi)$ is affected, with transformations given by Wigner D-matrices~\cite{Varshalovich1988}.
\begin{align}
    \Psi_{l,m}^{(n)}(\omega; \mathcal{R}^{-1}(\alpha, \beta,\gamma)\mathbf{r}) = \sum_{m'=-l}^l D^l_{m,m'}(\alpha, \beta,\gamma) \Psi_{l,m'}^{(n)}(\omega; \mathbf{r})\,.
\end{align}
For the LVSWs, we must also apply the matrix $\mathcal{R}(\alpha, \beta,\gamma)$ to the left-hand side of the equation. Please note that the inverse operator $\mathcal{R}^{-1}(\alpha, \beta,\gamma) = \mathcal{R}(\gamma, \beta, \alpha)$ transforms the radius-vector $\mathbf{r}$. The result of applying the rotation by an angle $\beta$ about the $z$-axis to the SCW is
\begin{align}
    \Psi_{k_z, m}^{(n)}(\omega; \mathcal{R}^{-1}(0, \beta,0)\mathbf{r}) = \mathrm{e}^{-\mathrm{i}m\beta}  \Psi_{k_z, m}^{(n)}(\omega;\mathbf{r})\,,
\end{align}
and to the SPW is
\begin{align}
    \Psi_{\mathbf{k}}(\omega; \mathcal{R}^{-1}(0, \beta,0)\mathbf{r}) =
     \Psi_{\mathcal{R}(0, \beta,0)\mathbf{k}}(\omega; \mathbf{r})\,.
\end{align}

Translations of SSWs and SCWs in a local basis are described by block-diagonal matrices $\bm{\mathcal{C}}^{(1)}$. For SPWs, translation by $(-\mathbf{r}')$ introduces a phase factor:
\begin{align}
\label{eq_pw_translation}
    \Psi_{\mathbf{k}}(\omega; \mathbf{r} - \mathbf{r}') = \mathrm{e}^{-\mathrm{i}\mathbf{k} \cdot \mathbf{r}'}\Psi_{\mathbf{k}}(\omega; \mathbf{r})\,.
\end{align}
This relation, combined with Eqs.~\eqref{eq_pw_into_sw} and~\eqref{eq_pw_into_cw}, enables expansions of plane waves in SSW and SCW bases with respect to arbitrary reference points. Please note that the translation and expansion coefficients of longitudinal vector waves coincide with those of scalar waves, since the gradient operator is invariant under translations~\cite{Kim2004}.

Finally, permutations of Cartesian axes are useful when applying the \textit{S}-matrix formalism to 1D lattices in the SCW basis. For example, a lattice in the $xz$-plane can be mapped to the $xy$-plane via $(x,y,z) \to (z,x,y)$ without altering the physical state, since $\Psi_{\mathbf{k}}(\omega;\mathbf{r}) = \Psi_{\mathbf{k}'}(\omega;\mathbf{r}')$ and $\mathbf{L}_{\mathbf{k}}(\omega;\mathbf{r}) = \mathbf{L}_{\mathbf{k}'}(\omega;\mathbf{r}')$.

\section{Structure of the program and examples}\label{sec:structure}
This section outlines the main classes and methods implemented in \textit{acoustotreams} for modeling acoustic wave scattering and presents several examples illustrating how to use this functionality in practical simulations. To run the examples below, the following preamble is required:
\begin{lstlisting}[label={lst:import}]
>>> import acoustotreams as at
>>> import numpy as np
\end{lstlisting}
Additional examples, including full scripts, are available at \href{https://nikustimenko.github.io/acoustotreams}{https://nikustimenko.github.io/acoustotreams}. These examples are automatically re-executed whenever a new version of the program is released. 

\subsection{General structure}
The program \textit{acoustotreams} builds upon the framework of the Python package \textit{treams}, adapting it to acoustic scattering problems. A comprehensive description of \textit{treams} can be found in Ref.~\cite{Beutel2024Apr} and Chapter~4 of Ref.~\cite{Beutel2024thesis}. While the overall structure of both packages is similar, important differences arise from the underlying physical formulations: \textit{acoustotreams} employs scalar (and longitudinal) waves as basis functions, whereas \textit{treams} employs transverse vector waves for electromagnetic problems. This distinction affects both class definitions and the associated mathematical operations.

\begin{figure}[h!]
\centering
\includegraphics[width=\linewidth]{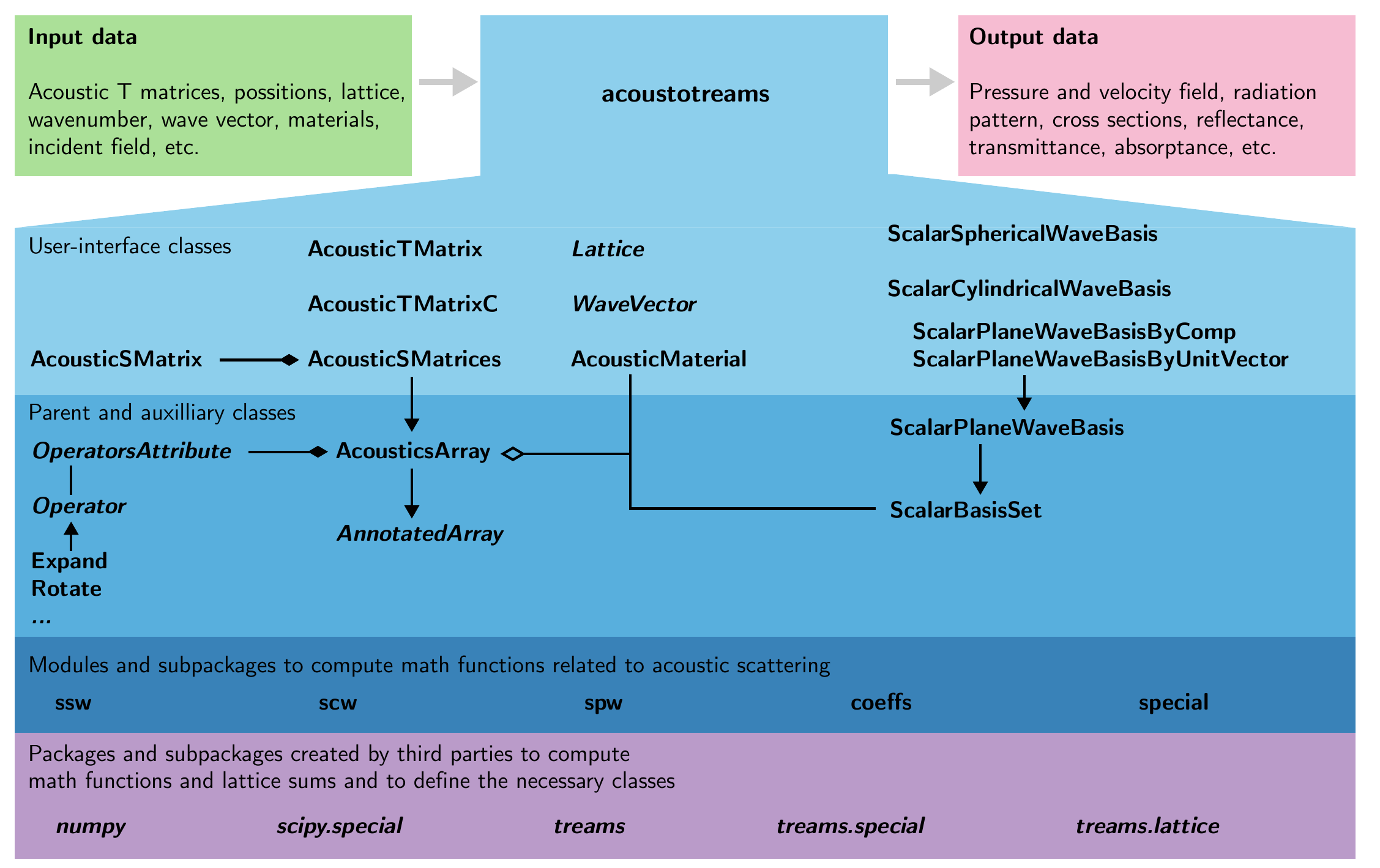}
\caption{Schematic representation of the structure and classes of \textit{acoustotreams}, including examples of input and output data. For a given finite or infinite arrangement of scatterers and an incident wave, the program computes the acoustic response using the \textit{T}-matrix method. The \textit{T}- and \textit{S}-matrix classes are derived from \texttt{AcousticsArray}, an annotated \textit{numpy} array that stores attributes such as wavenumber, background material, basis set, etc. Basis sets of scalar waves are implemented as subclasses of \texttt{ScalarBasisSet}. For both \texttt{AcousticsArray} and \texttt{ScalarBasisSet}, unitary operators are available as attributes (or static methods). The computations rely on both internal scattering functions and external mathematical libraries. Imported components are shown in italics. Arrows with filled triangles, filled diamonds, open diamonds, and plain lines denote inheritance, composition, aggregation, and association, respectively (UML notation).}
\label{fig:scheme}
\end{figure}

The structure of \textit{acoustotreams} is illustrated in Fig.~\ref{fig:scheme}, which presents the main classes as a Unified Modeling Language (UML) diagram. The package relies on a range of mathematical functions, including general-purpose routines from \textit{numpy} and \texttt{scipy.special}, as well as scattering-specific functions from \texttt{treams.special}. The functions for computing lattice sums are imported from \texttt{treams.lattice}.

As discussed in Sec.~\ref{sec:theory}, three scalar basis sets--spherical, cylindrical, and plane waves--are implemented, together with their longitudinal vector counterparts and the far-field amplitudes of singular waves. Since the external modules do not provide these functions, they are implemented directly in the subpackage \texttt{special} of \textit{acoustotreams}.

Translation and rotation coefficients, as well as basis transformation coefficients between SSWs, SCWs, and SPWs, are implemented as vector functions in the modules \texttt{ssw}, \texttt{scw}, and \texttt{spw}. The module \texttt{coeffs} contains functions for computing reflection and transmission coefficients at planar interfaces (Fresnel coefficients), as well as \textit{T}-matrix coefficients (Mie coefficients) for multilayered spheres in the SSW basis and infinite multilayered cylinders in the SCW basis.

\subsection{The class \texttt{AcousticMaterial}}
Materials in \textit{acoustotreams} are represented by instances of the class \texttt{AcousticMaterial}, defined by three scalar parameters: \texttt{rho}, \texttt{c}, and \texttt{ct}, corresponding to the mass density $\rho$ [kg/m$^3$], the speed of sound (pressure waves) $c$ [m/s], and the speed of shear (transverse) waves $c^{\rm t}$ [m/s], respectively:
\begin{lstlisting}[label={lst:mat}] 
>>> mat = at.AcousticMaterial(rho=2230, c=5661, ct=3392) 
\end{lstlisting}
Instead of $c^{\rm t}$, one may alternatively specify the Poisson ratio. A material can also be defined by its density and Lamé parameters. By default, the background material is air, with $\rho = 1.3$ kg/m$^3$ and $c = 343$ m/s.

\subsection{The class \texttt{AcousticsArray}}
User interaction with the program is primarily handled through ``physics-aware'' classes derived from \texttt{AcousticsArray}. This class extends \texttt{PhysicsArray} from \textit{treams}, which is essentially a \textit{numpy} array augmented with annotations stored in a dictionary (\texttt{AnnotationDict}). These annotations track physical parameters and define operators acting on the arrays.

Annotations may be assigned globally or per array dimension. In the latter case, compatibility is checked during binary operations: mismatched annotations trigger a warning. For elementwise operations, annotations are compared along each dimension, while for matrix multiplication, they are checked between the last dimension of the first operand and the first dimension of the second.

The attributes (annotations) of an \texttt{AcousticsArray} include the angular wavenumber in air $k_0 = \omega/c_0$ (with $c_0 = 343$ m/s), the background material, the basis set, and the mode type (see Table~\ref{tab:attributes}). For SSWs and SCWs, the mode type distinguishes between regular and singular modes. For SPWs, defined via the transverse wave vector $\mathbf{k}_{\parallel}$, the mode type is either \texttt{up} or \texttt{down}, indicating the propagation direction $\uparrow$ or $\downarrow$. For periodic systems, additional attributes include the lattice and the wave vector $\mathbf{k}_{\parallel}$, represented by instances of \texttt{Lattice} and \texttt{WaveVector}.

\begin{table}[]
    \begin{tabular}{l l l}
         \hline \\
         Description& Attribute & Type\\
         \hline \\
         Angular wavenumber in the air& \texttt{k0} & \texttt{float} \\
         Background material& \texttt{material} & \texttt{AcousticMaterial} \\
         Basis set & \texttt{basis} & \texttt{ScalarBasisSet} \\
         Mode type (depending on basis set)& \texttt{modetype} & \texttt{str} \\
         Lattice& \texttt{lattice} & \texttt{Lattice} \\
         Wave vector (along the lattice dimension)& \texttt{kpar} & \texttt{WaveVector} \\
         \hline
    \end{tabular}
    \caption{Attributes of~\texttt{AcousticsArray}.}
    \label{tab:attributes}
\end{table}

\subsection{Scalar basis sets}
As discussed in Sec.~\ref{sec:basis}, \textit{acoustotreams} provides three scalar wave bases for solving Eqs.~\eqref{eq_p_helmholtz} and~\eqref{eq_v_helmholtz} in different coordinate systems. In \textit{acoustotreams}, choosing an appropriate basis can significantly simplify a given problem. All basis sets are implemented as subclasses of \texttt{ScalarBasisSet}. In the following, we consider the most convenient methods for creating basis sets in \textit{acoustotreams}.

SSWs are indexed by degree $l \in \mathbb{N}_0$ and order $m \in \{ -l,-l+1,...,l\}$. A truncated basis of SSWs at a maximum multipole degree $l_{\rm max}$ can be constructed as follows:
\begin{lstlisting}[label={lst:ssw}]
>>> at.ScalarSphericalWaveBasis.default(1, 2, [[0, 0, 0], [0, 0, 1]])
ScalarSphericalWaveBasis(
    pidx=[0 0 0 0 1 1 1 1],
    l=[0 1 1 1 0 1 1 1],
    m=[0 -1  0  1  0 -1  0  1],
    positions=[[0. 0. 0.], [0. 0. 1.]],
)
\end{lstlisting}
where we have created an SSW basis up to the dipolar degree $l_{\rm max} = 1$. The degree can then be either 0 or 1, while the order $m$ is 0 for $l = 0$ and takes three values ($-1, 0, 1$) for $l = 1$. Here, the additional fields \texttt{pidx} and \texttt{positions} enable expansions with respect to multiple centers. By default, a single expansion center at $\mathbf{r} = \mathbf{0}$ is assumed.

Similarly, an SCW basis indexed by $k_z \in \mathbb{R}$ and $m \in \mathbb{Z}$ can be defined as:
\begin{lstlisting}[label={lst:scw}]
>>> at.ScalarCylindricalWaveBasis.default([-1, 1], 1)
ScalarCylindricalWaveBasis(
    pidx=[0 0 0 0 0 0],
    kz=[-1. -1. -1.  1.  1.  1.],
    m=[-1  0  1 -1  0  1],
    positions=[[0. 0. 0.]],
)
\end{lstlisting}
Since we can conveniently describe a 1D periodic lattice along the $z$-axis in the SCW basis, we can create a basis that describes diffraction orders with $ \left( k_{\parallel} + G \right)$ as
\begin{lstlisting}[label={lst:scw_diff}]
>>> at.ScalarCylindricalWaveBasis.diffr_orders(1, 1, lattice=2*np.pi, bmax=1.05)
ScalarCylindricalWaveBasis(
    pidx=[0 0 0 0 0 0 0 0 0],
    kz=[0. 0. 0. 1. 1. 1. 2. 2. 2.],
    m=[-1  0  1 -1  0  1 -1  0  1],
    positions=[[0. 0. 0.]],
)
\end{lstlisting}
where \texttt{bmax} sets the cutoff length for reciprocal lattice vectors $G \leq b_{\rm max}$. For 1D lattices, the argument \texttt{lattice} can accept a float number, which determines the lattice constant, as well as an instance of the class \texttt{Lattice}, being \texttt{at.Lattice(2*np.pi)} in our case.

The SPW basis differs in that it is defined with respect to a single center and can be constructed either via a unit vector $\hat{\mathbf{k}}$ multiplied by the wavenumber (the class \texttt{ScalarPlaneWaveBasisByUnitVector}) or two components $\mathbf{k}_{\parallel}$ (the class \texttt{ScalarPlaneWaveBasisByComp}). In the latter case, the remaining component ($z$ by default) is determined from the dispersion relation, with its sign specified by \texttt{modetype}, which can be either \texttt{"up"} (by default) or \texttt{"down"}.
Moreover, we can transform these basis sets into one another. 

Similarly to SCWs, we can use the following method  of the class \texttt{ScalarPlaneWaveBasisByComp} to generate diffraction orders for a 2D lattice in the $xy$-plane,
\begin{lstlisting}[label={lst:spw_diff}]
>>> lattice = at.Lattice.square(2 * np.pi)
>>> at.ScalarPlaneWaveBasisByComp.diffr_orders([0, 0], lattice, bmax=1))
ScalarPlaneWaveBasisByComp(
    kx=[ 0.  0.  0.  1. -1.],
    ky=[ 0.  1. -1.  0.  0.],
)
\end{lstlisting}

\subsection{Operators}
A distinctive feature of \textit{acoustotreams} is its explicit support for a wide range of operators discussed in Sec.~\ref{sec:theory} and summarized in Table~\ref{tab:operators}. First, the \textit{T}- and \textit{S}-matrices are operators that transform the incident field expansion into the scattered field expansion in a multipolar basis or the incoming field expansion into the outgoing field expansion in a plane-wave basis, respectively. Moreover, the implemented operators include the expansion of regular waves in another basis given by Eqs.~\eqref{eq_pw_into_sw}-\eqref{eq_cw_into_sw}; the expansion of singular (scattered) multipolar waves in regular (incident) multipolar waves $\bm{\mathcal{C}}^{(3)}$ given by Eq.~\eqref{eq_transl_sing}; the expansion of multipolar waves in those of the same type $\bm{\mathcal{C}}^{(1)}$ given by Eqs.~\eqref{eq_transl_reg}; the expansion of singular multipolar waves in regular multipole waves for a lattice $\bm{\Sigma}$ given by Eq.~\eqref{eq_lattice_sums_def}; and the expansion of singular waves in a lattice in another basis given by Eqs.~\eqref{eq_sw_into_cw_lattice}-\eqref{eq_cw_into_pw_lattice}. The other operators are also rotations, translations, permutations of Cartesian axes, operators to evaluate the pressure $p(\omega;\mathbf{r})$ and velocity $\mathbf{v}(\omega;\mathbf{r})$ fields, and also an operator to evaluate the far-field amplitude of the scattered pressure field $p_{\rm FF}(\omega;\mathbf{n})$ (see~\ref{sec:radpattern}). 

\begin{table}[h!]
    \begin{tabular}{l l}
         \hline \\
         Operator name& Description\\
         \hline \\
         \texttt{Rotate} & Rotate by Euler angles \\
         \texttt{Translate} & Translate by a vector \\
         \texttt{Expand} & Expand in another basis/at other positions \\
         \texttt{ExpandLattice} & Expand in another basis/at other positions  assuming a periodic lattice \\
         \texttt{Permute} & Permute the axes of SPWs \\
         \texttt{PField, VField} & Evaluate the field at specified positions \\
         \texttt{PAmplitudeFF} & Evaluate the far-field amplitude of the scattered field in specified directions \\
         \hline
    \end{tabular}
    \caption{Operators in \textit{acoustotreams}.}
    \label{tab:operators}
\end{table}

Operators are implemented via the class \texttt{Operator}, imported from \textit{treams} without any changes. However, its subclasses in Table~\ref{tab:operators} were changed accordingly for the scalar basis sets. Let us consider rotations as an example (see Sec.~\ref{sec:other}) and define a rotation by angle $\pi/2$ about the $z$-axis as an instance of the class \texttt{Rotate}:
\begin{lstlisting}[label={lst:rotate}]
>>> rot = at.Rotate(np.pi / 2)
\end{lstlisting}
Its matrix representation in a given basis can be obtained by:
\begin{lstlisting}[label={lst:rotate2}]
>>> swb = ScalarSphericalWaveBasis.default(1)
>>> rot(basis=swb)
AcousticsArray(
...
\end{lstlisting}
Moreover, this operator can also act directly on an array if its basis is specified. A convenient way to specify the basis provided by \textit{acoustotreams} is to use instances of the class \texttt{AcousticsArray},
\begin{lstlisting}[label={lst:rotate3}]
>>> arr = AcousticsArray(np.eye(4), basis=swb)
>>> rot @ arr @ rot.inv
AcousticsArray(
...
\end{lstlisting}
Please note that the size of the array must be equal to the length of its basis. 

For convenience, operators are wrapped as the class \texttt{OperatorAttribute} for instances of the class \texttt{AcousticsArray}. In this case, the operator can be applied simply as \texttt{arr.rotate(np.pi / 2)}, which is equivalent to \texttt{rot {@} arr} for 1D arrays and operators without the inverse; otherwise, it corresponds to \texttt{rot {@} arr {@} rot.inv} as follows:
\begin{lstlisting}[label={lst:rotate4}]
>>> (rot @ arr @ rot.inv == arr.rotate(np.pi / 2)).all()
True
\end{lstlisting}

Notably, operators in \textit{acoustotreams} are not restricted to \texttt{AcousticsArray} instances and can be applied to instances of custom Python classes, provided the required attributes are defined. 

\subsection{Acoustic \textit{T}-matrix}\label{sec:tmatrix}
\begin{figure}[h!]
    \centering
    \includegraphics[width=\linewidth]{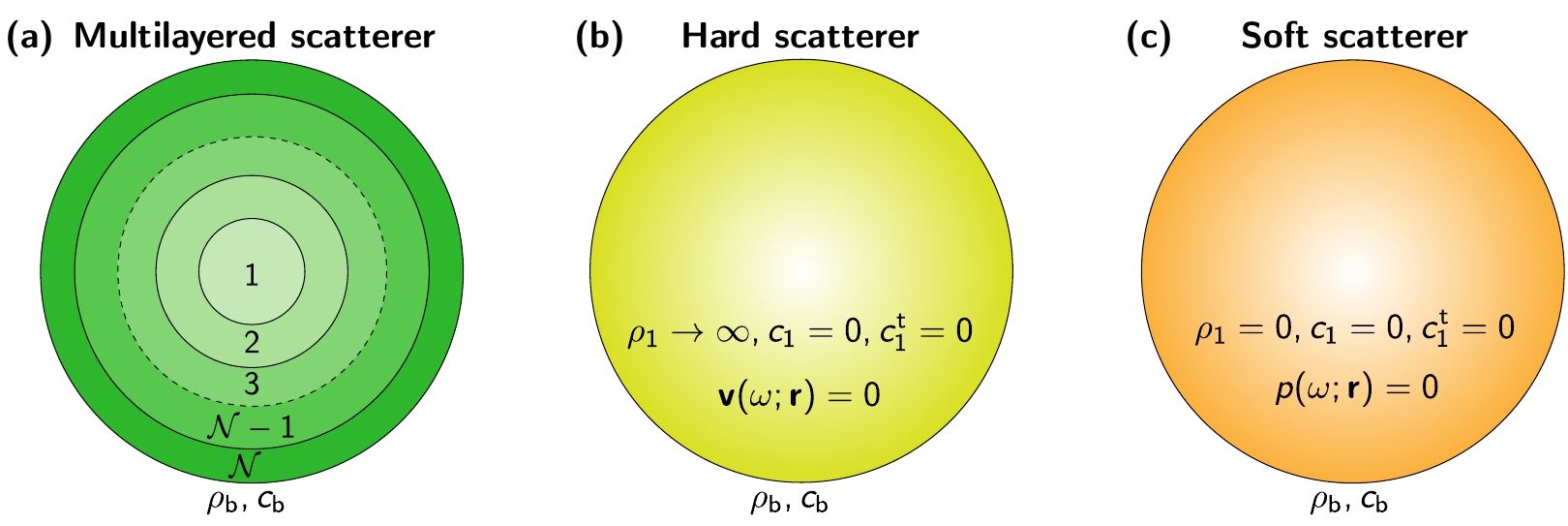}
    \caption{Scatterers, for which the acoustic \textit{T}-matrix can be computed semi-analytically in \textit{acoustotreams}, include (a) $\mathcal{N}$ concentric layers of materials, defined by their material parameters $(\rho_i, c_i, c_i^{\rm t})$ and objects with (b) hard and (c) soft interface conditions in a fluid background medium with $(\rho_{\rm b},c_{\rm b})$. The interface can be spherical or cylindrical. In the latter case, $c_i^{\rm t} \equiv 0$ for any material ($i$) in panel (a).}
    \label{fig:bc}
\end{figure}

To simulate the scattering of acoustic waves by individual objects and their arrangements in \textit{acoustotreams}, their acoustic \textit{T}-matrices are required. The classes \texttt{AcousticTMatrix} and \texttt{AcousticTMatrixC}, both derived from \texttt{AcousticsArray}, implement acoustic \textit{T}-matrices in the SSW and SCW bases, respectively.
On the one hand, these classes impose additional constraints on attributes and parameters. For instance, \texttt{AcousticTMatrix} requires the basis to be an instance of \texttt{ScalarSphericalWaveBasis}, and the mode type is automatically converted from \texttt{regular} to \texttt{singular}.
On the other hand, they provide convenient functionality. For example, the method \texttt{AcousticTMatrix.sphere} computes the acoustic \textit{T}-matrix of a spherical scatterer--either homogeneous or composed of concentric layers--using a semi-analytical approach. Such a computation is achieved by enforcing interface conditions at material interfaces, which depend on the corresponding material parameters. In \textit{acoustotreams}, the following spherical interfaces are supported (see Fig.~\ref{fig:bc}): fluid–fluid, fluid–solid, solid–fluid, solid–solid, soft–fluid, and hard–fluid.
Please note that (1) the background material must be a fluid with real-valued parameters, and (2) hard and soft scatterers are restricted to a single layer.

\subsubsection{Single scatterers}\label{sec:tmatrix_single}
As an example, we consider a multilayered sphere consisting of a fluid core and two solid shells [$\mathcal{N} = 3$ in Fig.~\ref{fig:bc}(a)]. In \textit{acoustotreams}, scatterers may be absorptive, i.e., they can have complex material parameters. We define the materials, compute the \textit{T}-matrix of the sphere at frequency $\omega/(2\pi) = 9.1$ kHz, and finally evaluate the rotation-averaged scattering and extinction cross sections~\eqref{eq_cross_avg} (line 12 of the following code):
\begin{lstlisting}[label={lst:sphere}]
>>> mat_1 = at.AcousticMaterial(rho=1050+50j, c=2350-1100j)
>>> mat_2 = at.AcousticMaterial(rho=2331+100j, c=8490-1400j, ct=5660-500j) 
>>> mat_3 = at.AcousticMaterial(rho=2230+80j, c=5661-1200j, ct=3392-1000j) 
>>> mat_b = at.AcousticMaterial(rho=1000, c=np.sqrt(21)*100)
>>> sphere = at.AcousticTMatrix.sphere(
...     lmax=4,
...     k0=2*np.pi*9100/343,
...     radii=[.005, .015, .02],
...     materials=[mat_1, mat_2, mat_3, mat_b]
...     ) 
>>> sphere.xs_sca_avg, sphere.xs_ext_avg
(1.3589e-03, 1.3782e-03)
\end{lstlisting}
Please note that these quantities can only be computed in the global basis.

Next, we consider a nonspherical scatterer--a cone--for which the \textit{T}-matrix ($l_{\rm max} = 5$) in the frequency range 5-25.5 kHz (206 points) has been computed using the FEM model~\cite{tm_comsol} and stored in the file \texttt{cone.hdf5}. The material and geometrical parameters were taken from Ref.~\cite{Ustimenko2025Apr}. After loading the data and constructing an \texttt{AcousticTMatrix} instance with the required attributes in Table~\ref{tab:attributes}, except \texttt{modetype}, we compute the scattering cross section for a given incident wave [Eq.~\eqref{eq_sigma_sca}] in line 13 of the code:
\begin{lstlisting}[label={lst:cone}]
>>> import acoustotreams as at
>>> import numpy as np
>>> import h5py

>>> file = h5py.File("cone.hdf5", "r")
>>> tm_set = file["tmatrix"][()]

>>> mat_b = at.AcousticMaterial(rho=1000, c=np.sqrt(21)*100)
>>> k0 = 2*np.pi*9100/343
>>> cone = at.AcousticTMatrix(tm_set[41, :, :], k0=k0, material=mat_b)
>>> inc = at.plane_wave_scalar([0, 0, k0], k0=k0, material=mat_b)
>>> cone.xs(inc)[0]
(2.1485e-05)
\end{lstlisting}
We can also compute the expansion coefficients of the scattered field~\eqref{eq_T_mat} and use them to evaluate the pressure field via Eq.~\eqref{eq_p_expansion} using the \texttt{pfield} operator:
\begin{lstlisting}[label={lst:cone2}]
>>> sca = cone.sca(inc)
>>> sca.pfield([-0.005, 0, 0.01])
[-0.02231387+0.1590198j]
\end{lstlisting}
To verify the condition $r > R_{\rm c}$, the property \texttt{AcousticTMatrix.valid\_points} is provided; it returns \texttt{True} if $r > R_{\rm c}$ and \texttt{False} otherwise.

\subsubsection{Multiple scattering}\label{sec:tmatrix_multiple}
We now consider a system of multiple scatterers, such as a dimer composed of the cone and a homogeneous sphere arranged on a 1D lattice along the $z$-axis. The sphere has a radius of 0.65~cm and material parameters $\rho_1=7000+\mathrm{i}150$ kg/m$^3$, $c_1 = 100-\mathrm{i}10$ m/s, and $c^{\rm t}_1 = 30-\mathrm{i}10$ m/s. The cone remains the same.

Given the \textit{T}-matrices ($l_{\rm max} = 5$) and positions of the individual scatterers, we construct the block-diagonal matrix $\mathbf{T}^{\rm diag}$ using \texttt{AcousticTMatrix.cluster}. Matrices~\eqref{eq_T_local} and~\eqref{eq_Teff} in a local basis are then obtained using \texttt{interaction.solve()} without periodic boundary conditions and \texttt{latticeinteraction.solve()} with them, respectively. In the latter case, we obtain the effective \textit{T}-matrix as
\begin{lstlisting}[label={lst:chain}]
>>> positions = [
...     [-0.0085, 0, -0.0075], 
...     [0.0085, 0, 0.0075]
...     ]
>>> kz = 0.1 * k0
>>> lattice = at.Lattice(0.035)
>>> tm_diag = at.AcousticTMatrix.cluster([sphere, cone], positions)
>>> tm_eff = tm_diag.latticeinteraction.solve(lattice, kz * n)
\end{lstlisting}
where \texttt{n} denotes an acoustic analog of the refractive index for a background medium different from air, $n = 343\text{[m/s]}/c_{\rm b}$[m/s], and \texttt{k0} corresponds to a frequency of 17.5~kHz in air. As in the single-scatterer case, scattered fields can be computed via expansion coefficients, now including an additional lattice re-expansion [Eq.~\eqref{eq_lattice_reexpansion}]:
\begin{lstlisting}[label={lst:chain_field}]
>>> inc = at.plane_wave_scalar(
...     [np.sqrt(k0 * k0 - kz * kz), 0, kz], 
...     k0=k0, 
...     material=mat_b)
>>> sca = tm_eff.sca(inc)
>>> r = [0.02, 0, -0.01]
>>> swb = at.ScalarSphericalWaveBasis.default(0, positions=[r])
>>> sca.expandlattice(basis=swb).pfield(r)
[0.557+1.261j]
\end{lstlisting}

As noted above, SCWs provide a more efficient treatment of 1D lattices. Therefore, we convert the effective \textit{T}-matrix in the SSW basis into the SCW basis that corresponds to diffraction orders using Eq.~\eqref{eq_sw_into_cw_lattice}, and then compute the expansion coefficients:
\begin{lstlisting}[label={lst:chain_tmc}]
>>> cwb = at.ScalarCylindricalWaveBasis.diffr_orders(
...     kz=kz * n, 
...     mmax=5, 
...     lattice=lattice, 
...     bmax=4.1 * lattice.reciprocal, 
...     nmax=2, 
...     positions=positions)
>>> tmc = at.AcousticTMatrixC.from_array(tm_eff, cwb)
>>> sca_tmc = tmc.sca(inc)
\end{lstlisting}
In the SCW basis, lattice summation is already considered by the expansion coefficients in line 8, so no additional re-expansion is required to evaluate the field:
\begin{lstlisting}[label={lst:chain_tmc_field}]
>>> sca_tmc.pfield(r)
[0.557+1.261j]
\end{lstlisting}

The \texttt{AcousticTMatrixC} class is analogous to \texttt{AcousticTMatrix}, but uses the SCW basis implemented via \texttt{ScalarCylindricalWaveBasis}. For an infinite cylinder aligned along the $z$-axis, the \textit{T}-matrix can be computed with \texttt{AcousticTMatrixC.cylinder}. The cylinder may consist of multiple concentric layers, but the speed of shear waves must be zero in both the layers and the background medium. This limitation can be relaxed by introducing corresponding interface conditions on the cylindrical surface. Otherwise, cylindrical \textit{T}-matrices behave similarly to spherical ones.

\subsection{Acoustic \textit{S}-matrix}\label{sec:smatrix}

The classes \texttt{AcousticSMatrix} and \texttt{AcousticSMatrices}, also derived from \texttt{AcousticsArray}, implement the acoustic \textit{S}-matrix in the SPW basis defined by $\mathbf{k}_{\parallel}$ (diffraction orders). An \texttt{AcousticSMatrix} instance represents a single block from Eq.~\eqref{eq_pSa}, while \texttt{AcousticSMatrices} contains four such blocks (with consistent annotations), covering all combinations of incoming and outgoing propagation directions.
Table~\ref{tab:smat} summarizes the systems for which \textit{S}-matrix computations are implemented in \textit{acoustotreams}.

\begin{table}[h!]
\centering
    \begin{tabular}{l l}
         \hline \\
         Class method& Description\\
         \hline \\
         \texttt{interface} & Interface between two fluid media \\
         \texttt{propagation} & Propagation along a distance in a fluid medium \\
         \texttt{slab} & Slab of three fluid media \\
         \texttt{from\_array} & Periodic repetition of spherical or cylindrical \textit{T}-matrices \\
         \hline
    \end{tabular}
    \caption{The methods of the \texttt{AcousticSMatrices} class for the computation of acoustic \textit{S}-matrices.}
    \label{tab:smat}
\end{table}
 
In the following example, we consider a double-layer structure composed of identical 2D lattices, each unit cell containing the absorptive sphere and the cone from the previous example. First, we define $\mathbf{k}_{\parallel}$ and the lattice:
\begin{lstlisting}[label={lst:k0_lattice}]
>>> kpar = [0.1 * k0, 0]
>>> lattice = at.Lattice.square(period)
\end{lstlisting}
Next, we compute the effective \textit{T}-matrix and convert it to the \textit{S}-matrix using \texttt{from\_array}, with the SPW basis defined as:
\begin{lstlisting}[label={lst:sm}]
>>> tm_diag = at.AcousticTMatrix.cluster([sphere, cone], positions)
>>> tm_eff = tm_diag.latticeinteraction.solve(lattice, kpar * n)
>>> bmax = 3.1 * 2 * np.pi / period
>>> spwb = at.ScalarPlaneWaveBasisByComp.diffr_orders(kpar * n, lattice, bmax)
>>> sm = at.AcousticSMatrices.from_array(tm_eff, spwb)
\end{lstlisting}
where the value of \texttt{bmax} is chosen such that the SPW basis includes sufficiently many diffraction orders to accurately capture both the response of a single lattice and the near-field coupling between layers.
We then stack the \textit{S}-matrices along the $z$-axis using the \texttt{stack} method, which effectively computes their Redheffer star product~\cite{Redheffer1959}:
\begin{lstlisting}[label={lst:stack}]
>>> propagation = at.AcousticSMatrices.propagation(
...     r=[0, 0, 0.04], 
...     basis=pwb, 
...     k0=k0, 
...     material=mat_b
...     )
>>> stack = at.AcousticSMatrices.stack([sm, propagation, sm])
\end{lstlisting}
Finally, we evaluate the transmittance ($T$), reflectance ($R$), and absorptance ($1-T-R$) for a given incident wave using the \texttt{tr} method:
\begin{lstlisting}[label={lst:tr}]
>>> inc = at.plane_wave_scalar(
...     kpar * n, 
...     k0=k0, 
...     basis=pwb, 
...     material=mat_b, 
...     modetype="down")
>>> coeffs = stack.tr(inc)
>>> coeffs[0], coeffs[1], 1 - sum(coeffs)
0.10, 0.52, 0.38
\end{lstlisting}
where \texttt{modetype} specifies the propagation direction along the $z$-axis (default: \texttt{"up"}).

Additional useful methods of \texttt{AcousticSMatrices} include \texttt{periodic}, which constructs periodic repetitions of \textit{S}-matrices along the $z$-axis, and \texttt{band\_kz}, which computes the corresponding band structure.

\section{Validation}\label{sec:validation}
In this section, we validate the functionalities of \textit{acoustotreams} presented in Sec.~\ref{sec:structure}. First, we directly compare results obtained with \textit{acoustotreams} to those computed using an independent full-wave solver for acoustic scattering by single scatterers. Second, we verify the internal consistency of the program by comparing different descriptions of a periodic system with a complex unit cell using the various basis sets implemented in \textit{acoustotreams}.

In the following, we focus on validating the methods described in Sec.~\ref{sec:tmatrix}. The computation of reflectance, transmittance, and absorptance based on the \textit{S}-matrix formalism (Sec.~\ref{sec:smatrix}) has already been verified in Ref.~\cite{Ustimenko2026Jan}.

\subsection{Single scatterers}

\begin{figure}[h!]
    \centering
    \includegraphics[width=\linewidth]{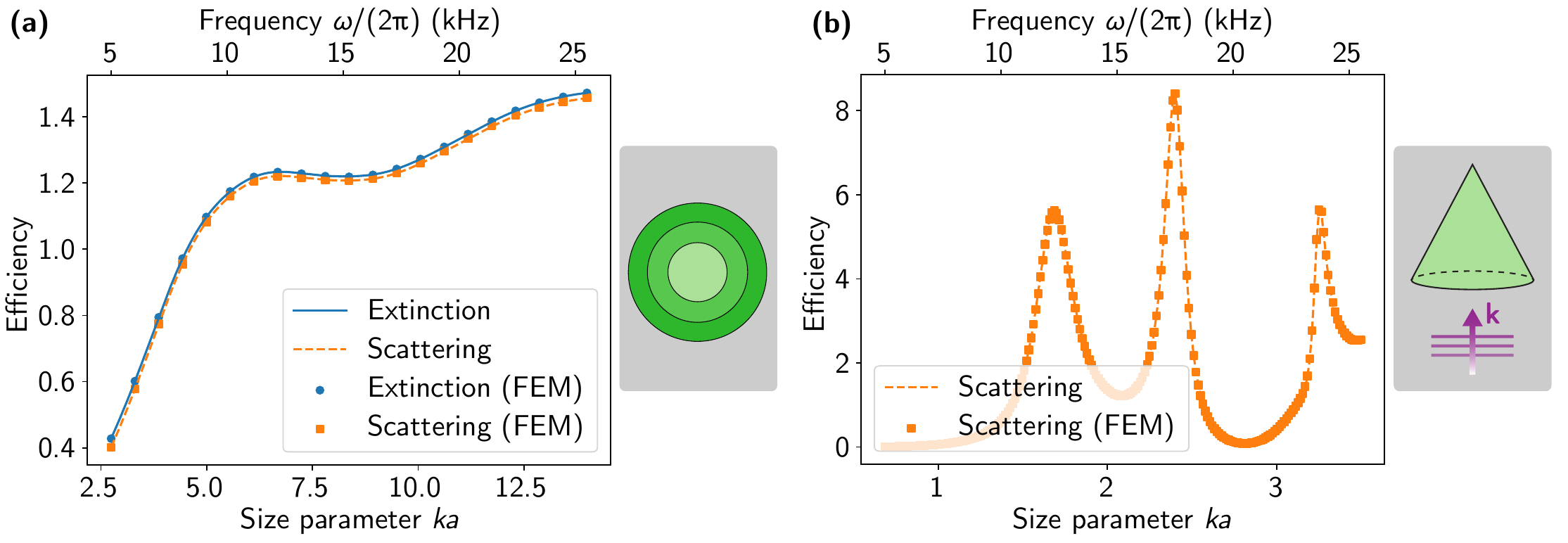}
    \caption{(a) Rotation-averaged extinction and scattering efficiencies of a sphere with three layers, shown on the right, computed using \textit{acoustotreams} (lines) and the FEM model in COMSOL Multiphysics\texttrademark, v.6.2~(markers). (b) Comparison of the scattering efficiency in Eq.~\eqref{eq_sigma_sca} for a homogeneous, nonabsorptive, fluid cone. The incident plane pressure wave is shown on the right, along with the cone-shaped scatterer. The material and geometric parameters are given in the corresponding paragraphs of Subsec.~\ref{sec:tmatrix_single}. The extinction and scattering efficiencies are defined as the extinction and scattering cross sections divided by $\pi a^2/4$, respectively. The geometrical size $a$ is equal to the sphere diameter of 4 cm in panel (a) and to the diameter of the cone base in panel (b), being 1 cm. The size parameter is defined as $ka$, where $k = \omega/c_{\rm b}$ is the wavenumber in the background medium.}
    \label{fig:4}
\end{figure}

Figure~\ref{fig:4} compares the extinction and scattering efficiencies for the spherical and conical scatterers introduced in Sec.~\ref{sec:tmatrix_single}, which we computed using two different approaches, namely \textit{acoustotreams} and the FEM model~\cite{tm_comsol}. Here, the extinction and scattering efficiencies are defined as the corresponding cross section divided by the geometrical cross section of a scatterer $\pi a^2/4$, where $a$ = 4 cm for the sphere and 1 cm for the cone. Within the FEM model, the extinction and scattering cross sections are evaluated as surface integrals of $\langle \mathbf{S}_{\rm tot} \rangle \cdot \mathbf{n}$ and $\langle \mathbf{S}_{\rm sca} \rangle \cdot \mathbf{n}$ over a spherical surface with a radius of $r > R_{\rm c}$, respectively, divided by the intensity of the incident field $I_0 = |p_0|^2/(2 \rho_{\rm b} c_{\rm b})$. Here, $\mathbf{S}_{\rm tot}$ and $\mathbf{S}_{\rm sca}$ denote the time-averaged energy flux vectors of the total and scattered fields, and $\mathbf{n}$ is the outward unit normal (see~\ref{sec:radpattern}). For both scatterers, the incident field was chosen as a plane-wave pressure field with amplitude $p_0 = 1$~Pa, propagating along the $z$-axis.

Figure~\ref{fig:4}(a) shows the rotation-averaged extinction and scattering efficiencies for the multilayered sphere as a function of frequency. These were obtained by first computing the acoustic \textit{T}-matrix in \textit{acoustotreams} with a maximum multipole degree $l_{\rm max} = 8$, and then evaluating the corresponding cross sections, as described in Sec.~\ref{sec:tmatrix_single}, and dividing them by $\pi a^2/4$. The scattering efficiency is slightly smaller than the extinction efficiency due to absorption within the scatterer. For spherical scatterers, the cross sections are independent of the incident direction and therefore coincide with their rotation-averaged values.

Figure~\ref{fig:4}(b) presents the scattering efficiency for the cone. Since absorption is absent, the extinction and scattering cross sections are equal. After computing the \textit{T}-matrix of the cone using the FEM model for axisymmetric scatterers with $l_{\rm max} = 5$, we obtained the scattering cross section of a plane pressure wave that impinges on the scatterer as shown in Fig.~\ref{fig:4}(b). 

Overall, Fig.~\ref{fig:4} demonstrates excellent agreement between \textit{acoustotreams} and the FEM results in both cases. We also emphasize that the cross sections of the sphere and the cone computed by \textit{acoustotreams} converge to the corresponding FEM values in the entire frequency range for $\ell_{\rm max} = 8$ and $\ell_{\rm max} = 3$, thereby confirming the rule $\ell_{\rm max} \approx \frac{1}{2} \left( k a \right) + 1$ for single scatterers in a homogeneous environment mentioned in Sec.~\ref{sec:multipolar_expansion}. 

\subsection{Lattice with a complex unit cell}\label{sec:validation_lattice}

Next, we consider a multiple-scattering problem involving a 1D lattice along the $z$-axis (see Sec.~\ref{sec:tmatrix_multiple}). The unit cell consists of the sphere and the cone located at $\mathbf{r}_1 = \left(-8.5,0,-7.5 \right)$ mm and $\mathbf{r}_2 = \left(8.5,0,7.5 \right)$ mm, respectively. The sphere has a radius of 6.5~mm, and its material parameters are listed at the beginning of Sec.~\ref{sec:tmatrix_multiple}.  The cone has a height and base diameter of 10~mm. The radii of the corresponding circumscribing spheres, centered at the centers of mass of the scatterers, are $R_{\rm c}^{(1)} = 6.5$~mm and $R_{\rm c}^{(2)} = 7.5$~mm. The lattice constant is $L = 35$~mm. The Bloch wavenumber is $k_{\parallel} = 0.1k_0$, where $k_0$ corresponds to a frequency of 17.5~kHz in air. The incident field is a plane-wave pressure field propagating along the $x$-axis. To ensure and also study the convergence of the results with $\ell_{\rm max}$, we recomputed the \textit{T}-matrix of the cone at 17.5 kHz up to $\ell_{\rm max} = 7$ using the same FEM model.

\begin{figure}[h!]
    \centering
    \includegraphics[width=\linewidth]{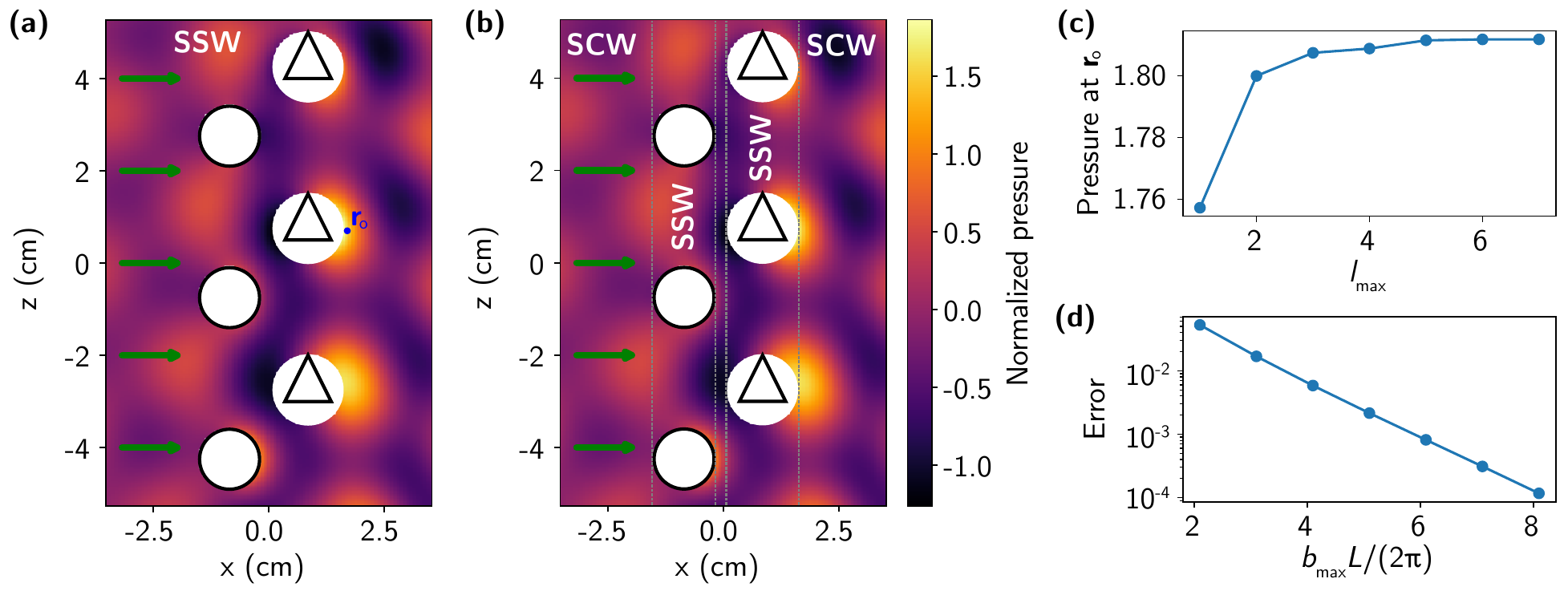}
    \caption{(a) Real part of the scattered pressure field in the SSW basis for a 1D lattice along the $z$-axis, whose unit cell contains spherical and cone-shaped scatterers. Green arrows indicate the incident plane pressure wave. (b) The same, but using the combined description based on the SSWs to compute the field inside the circumscribing cylinders (denoted by gray lines) and the SCWs to compute the field outside them. (c) The scattered pressure field value $|p_{\rm sph}|$ in the SSW basis at point $\mathbf{r}_{\mathrm{o}} = (16.5, 0, 7.5)$ mm. This point is indicated by a blue marker in panel (a). (d) Error of the SCW expansion defined by Eq.~\eqref{eq_error}.}
    \label{fig:5}
\end{figure}

Figure~\ref{fig:5}(a) shows the real part of the scattered pressure field computed in the SSW basis with maximum multipole degree $l_{\rm max} = 7$. The field was evaluated within the convergence domain of the SSW expansion, i.e., where $|\mathbf{r} - \mathbf{r}_1| > R_{\rm c}^{(1)}$ and $|\mathbf{r} - \mathbf{r}_2| > R_{\rm c}^{(2)}$. Regions where this condition is violated are shown in white. The field was obtained at each point of interest via the lattice re-expansion in Eq.~\eqref{eq_lattice_reexpansion}, which converged rapidly due to the use of Ewald summation~\cite{Beutel2023Jan}. Although the computations were performed for $|z| < L/2$, the results were extended over three unit cells using Bloch’s theorem to enhance visualization.

Figure~\ref{fig:5}(b) demonstrates the transition between the SSW and SCW descriptions for a 1D lattice provided by Eq.~\eqref{eq_sw_into_cw_lattice}. The computational domain was divided into three subdomains (indicated by gray dashed lines), corresponding to non-overlapping circumscribing cylinders around the scatterers. Inside these cylinders and between the scatterers, the field was computed using SSWs, while the outside field was re-expanded in SCWs with $b_{\rm max}L/(2\pi) = 4.1$. The continuity of the field across domain boundaries confirms the correctness of the lattice-sum implementation for scalar waves. Moreover, combining SSW and SCW representations resulted in a computational speedup of approximately a factor of three for this system with $l_{\rm max} = 7$.

Figures~\ref{fig:5}(c) and~\ref{fig:5}(d) illustrate convergence properties. In Fig.~\ref{fig:5}(c), the magnitude of the pressure field $|p_{\rm sph}(\mathbf{r}_{\mathrm{o}})|$ is evaluated at the probe point $\mathbf{r}_{\mathrm{o}} = (16.5, 0, 7.5)$~mm for increasing values of $l_{\rm max}$. The results converge at $l_{\rm max} = 5$. 

In the SCW description, the expansion formally involves an infinite number of diffraction orders [Eq.~\eqref{eq_sw_into_cw_lattice}]. In practice, the series is truncated using a cutoff parameter $b_{\rm max}$, neglecting higher-order evanescent modes that primarily contribute in the near-field region. Figure~\ref{fig:5}(d) shows the error defined as the maximum of the absolute difference over the validity domain of the SCW expansion, i.e.,
\begin{align}
\label{eq_error}
    \text{Error} = \max_{\mathbf{r} \in \text{domain}} \left| \Re[p_{\rm sph}(\mathbf{r})] - \Re[p_{\rm cyl}(\mathbf{r})]\right|\,,
\end{align}
where $p_{\rm cyl}(\mathbf{r})$ was computed for different values of $b_{\rm max}$. The error decreases as $b_{\rm max}$ increases, confirming the convergence of the SCW expansion.

A similar strategy can be applied to 2D lattices using SPW expansions [cf. Eq.~\eqref{eq_sw_into_pw_lattice}], which can further improve the computational efficiency~\cite{Beutel2024Apr}.

\section{Conclusion}
In this work, we presented a \textit{T}-matrix-based computational framework for acoustic multiple-scattering problems, implemented as a Python package called \textit{acoustotreams}. To extend the range of geometries that can be simulated and to maximize computational efficiency, three basis sets were implemented, namely scalar spherical, cylindrical, and plane waves. User interaction with the program is facilitated through dedicated classes that track physics-related attributes and provide methods implementing various wave operators, based on analytically known functions defined in the program’s submodules and subpackages. Once the \textit{T}-matrices of individual scatterers are known, the framework enables efficient computation of the acoustic response of individual scatterers, finite clusters, and infinite lattices with complex unit cells, relying in the latter case on rapidly converging lattice sums. The description of wave propagation in stratified media is also supported within the \textit{S}-matrix formalism.

The \textit{acoustotreams} package is open-source and available on GitHub under the MIT license, together with comprehensive online documentation and automated tests. Importantly, it can be installed on Linux, Windows, and macOS via the Python Package Index. This manuscript presents several illustrative examples of its use, while additional examples are provided in the documentation. The program has already been employed in Refs.~\cite{Ustimenko2025Apr,Ustimenko2026Jan, Ustimenko2026Feb, Demeulenaere2025Dec}, which offer further benchmarks and validation tests.

\section*{Acknowledgements}
The authors thank Dominik Beutel, Lukas Rebholz, Benedikt Zerulla, Jan David Fischbach, Oscar Demeulenaere, Athanasios Athanassiadis, and Peer Fischer for fruitful discussions.  The authors are also grateful to Ilya Deriy and Markus Nyman for their help with the FEM simulations. The authors acknowledge support through the Deutsche Forschungsgemeinschaft (DFG, German Research Foundation) under Germany’s Excellence Strategy via the Excellence Cluster 3D Matter Made to Order (EXC-2082/2, Grant No. 390761711) and from the Carl Zeiss Foundation via CZF-Focus@HEiKA.

\appendix

\section{Far-field amplitudes}\label{sec:radpattern}
The scattered pressure and velocity fields can be used to determine the radiation pattern of a single scatterer or a cluster of scatterers. The far-field scattered power $\mathrm{d}P_{\rm sca} = \langle \mathbf{S}_{\rm sca} \rangle \cdot \mathbf{n}\, \mathrm{d} A$ in the direction specified by the unit vector $\mathbf{n}$ is governed by the time-averaged energy flux of the scattered acoustic field, $\langle \mathbf{S}_{\rm sca} \rangle = \frac{1}{2} \, \Re\left[p_{\rm FF}\mathbf{v}^\ast_{\rm FF}\right]$~\cite{Williams1999}. From Eqs.~\eqref{eq_psca_ff_sph} and~\eqref{eq_psca_ff_cyl}, it follows that $\mathbf{v}_{\rm FF} \cdot \mathbf{n} \propto p_{\rm FF}$, implying $\langle \mathbf{S}_{\rm sca} \rangle \cdot \mathbf{n} \propto |p_{\rm FF}|^2$. Therefore, only the computation of $p_{\rm FF}$ is implemented in \textit{acoustotreams}.

\subsection{Spherical waves}
In spherical coordinates, the surface element scales as $\mathrm{d} A \propto r^2$, while the far-field amplitudes of the pressure and velocity fields, $p_{\rm FF}$ and $\mathbf{v}_{\rm FF}$, are given by
\begin{align}
    p_{\rm sca}(\omega; \mathbf{r})\big{|}_{kr \gg 1} = \frac{\mathrm{e}^{\mathrm{i}kr}}{r} p_{\rm FF} (\omega; \mathbf{n})\,, \quad \mathbf{v}_{\rm sca}(\omega; \mathbf{r})\big{|}_{kr \gg 1} = \frac{\mathrm{e}^{\mathrm{i}kr}}{r} v_{\rm FF}  (\omega; \mathbf{n})\mathbf{n}\,,
\end{align}
where $\mathbf{n} = \left( \sin \theta_\mathbf{r} \cos \varphi_\mathbf{r}, \sin \theta_\mathbf{r} \sin \varphi_\mathbf{r}, \cos \theta_\mathbf{r} \right)$ and $k \in \mathbb{R}_+$.

For a cluster of scatterers, these amplitudes in the SSW basis read as 
\begin{align}
\label{eq_psca_ff_sph}
    p_{\rm FF} (\omega; \mathbf{n}) = \sum_{i=1}^N \sum_{l,m} a_{i,l,m} \psi_{l,m}^{(3)}(\omega; \mathbf{n}, \mathbf{r}_i)\,, \quad v_{\rm FF} (\omega; \mathbf{n}) = \frac{1}{\rho_{\rm b}c_{\rm b}} p_{\rm FF} (\omega; \mathbf{n})\,,
\end{align}
where the far-field asymptotics of singular SSWs [Eq.~\eqref{eq_ssw_def} with $n = 3$] with respect to the expansion center at $\mathbf{r}_i =  (x_i, y_i, z_i)$ is~\cite{Ganesh2008Oct},
\begin{align}
\label{eq_ssw_ff_def}
    \psi_{l,m}^{(3)}(\omega; \mathbf{n} , \mathbf{r}_i) = \frac{\mathrm{e}^{-\mathrm{i}k(\mathbf{n}\cdot\mathbf{r}_i)}}{k} (-\mathrm{i})^{l + 1} Y_{l,m}(\theta_\mathbf{r}, \varphi_\mathbf{r})\,.
\end{align}

\subsection{Cylindrical waves}
In cylindrical coordinates, the surface element scales as $\mathrm{d} A \propto \rho$, and the far-field amplitudes are
\begin{align}
\label{eq_psca_ff_cyl}
\begin{aligned}
    p_{\rm sca}(\omega; \mathbf{r})\big{|}_{k_{\rho}\rho \gg 1} = \begin{cases}
        \frac{\mathrm{e}^{\mathrm{i}k_{\rho}\rho}}{\sqrt{\rho}} p_{\rm FF} (\omega; \mathbf{n})\,, & k_{\rho} \in \mathbb{R}\,,\\
        0,& k_{\rho} \in \mathbb{C}\,,
    \end{cases}  \quad 
    \mathbf{v}_{\rm sca}(\omega; \mathbf{r})\big{|}_{k_{\rho}\rho \gg 1} = 
    \begin{cases}
    \frac{\mathrm{e}^{\mathrm{i}k_{\rho}\rho}}{\sqrt{\rho}} v_{\rm FF}  (\omega; \mathbf{n})\mathbf{n}\,,& k_{\rho} \in \mathbb{R}\,,\\
    0,& k_{\rho} \in \mathbb{C}\,,
    \end{cases}
\end{aligned}
\end{align}
where $\mathbf{n} = \left( \cos \varphi_\mathbf{r}, \sin \varphi_\mathbf{r}, 0 \right)$. The asymptotic expression for $\mathbf{v}_{\rm sca}$ may also include a term proportional to $k_z \,\hat{\mathbf{z}}$, but it is omitted here since it does not contribute to the outgoing radiation.

For a cluster of scatterers, the amplitudes in Eq.~\eqref{eq_psca_ff_cyl} take the form of
\begin{align}
    p_{\rm FF} (\omega; \mathbf{n}) = \sum_{i=1}^N \sum_{k_z,m} a_{i,k_z,m} \psi_{k_z,m}^{(3)}(\omega; \mathbf{n}, \mathbf{r}_i)\,, \quad v_{\rm FF} (\omega; \mathbf{n}) = \frac{1}{\rho_{\rm b}c_{\rm b}} \sqrt{\frac{k_{\rho}}{k}}p_{\rm FF} (\omega; \mathbf{n})\,,
\end{align}
where the far-field asymptotics of singular SCWs [Eq.~\eqref{eq_scw_def} with $n = 3$] with respect to the expansion center at $\mathbf{r}_i = (x_i, y_i, 0)$ is~\cite{Ganesh2009},
\begin{align}
\label{eq_scw_ff_def}
    \psi_{k_z,m}(\omega;\mathbf{n},  \mathbf{r}_i) &= \sqrt{\frac{2}{\pi k_{\rho}}}(-\mathrm{i})^{|m|}\mathrm{e}^{-\mathrm{i}\pi/4}\mathrm{e}^{-\mathrm{i}k_{\rho}(\mathbf{n} \cdot \mathbf{r}_i)} \mathrm{e}^{\mathrm{i} m \varphi + \mathrm{i} k_z z}\,.
\end{align}

\section{Translation coefficients for scalar spherical and cylindrical waves}\label{translation_coeffs}
The translation coefficients for SSWs are given by
\begin{align}
\begin{aligned}
    \mathcal{C}_{l', m', l, m}^{(n)}(\omega; \mathbf{r}) &= (-1)^m \,\mathrm{i}^{l'-l} \sqrt{4\pi (2l+1)(2l'+1)} \\
    &\times \sum_q \mathrm{i}^{q} \sqrt{2q+1} \,  \begin{pmatrix}
            l& l' & q \\
            m & -m' & m'-m
        \end{pmatrix}
        \begin{pmatrix}
            l& l' & q \\
            0& 0& 0
        \end{pmatrix}\Psi^{(n)}_{q,m-m'}(\omega; \mathbf{r})\,,
\end{aligned}
\end{align}
where $q \in \left\{\ell'+\ell, \ell'+\ell-2, ..., \max\left(|\ell'-\ell|, |m' - m| \right) \right\}$, i.e., the sum index $q$ runs over all values where the Wigner 3j-symbols are nonzero~\cite{stein1961, sack1964, Danos1965May, Gonis2000}. 

The translation coefficients for SCWs are~\cite{Martin2006Aug} 
\begin{align}
    \mathcal{C}_{k_z', m', k_z, m}^{(n)}(\omega; \mathbf{r}) = 
    \begin{cases}
        Z^{(n)}_{m-m'}\left(\sqrt{k^2 - k_z^2} \rho_{\mathbf{r}} \right) \mathrm{e}^{\mathrm{i}(m - m') \varphi_{\mathbf{r}} + \mathrm{i} k_z z}\,,& k_z=k_z'\,, \\
        0\,,& k_z \neq k_z'\,.
    \end{cases}
\end{align}

The translation coefficients for longitudinal vector waves are identical, since the gradient operator is invariant under translations.

\section{The scattered velocity field of a lattice}\label{sec_lattice_field}
Table~\ref{tab:asymptotics} lists the regular SSWs and SCWs [Eqs.~\eqref{eq_ssw_def}-\eqref{eq_vsw_def} and Eqs.~\eqref{eq_scw_def}-\eqref{eq_vcw_def}, respectively] that remain finite at $\mathbf{r} = \mathbf{0}$.
\begin{table}[h!]
    \centering
    \begin{tabular}{|c|c|c|}
    \hline
        & Scalar & Longitudinal vector \\
        \hline
         \text{Spherical}&  $\Psi^{(1)}_{0,0} = \dfrac{1}{\sqrt{ 4\pi}}$& $\mathbf{L}^{(1)}_{1,m} = \dfrac{1}{\sqrt{12\pi}} \times
         \begin{cases}
         \frac{1}{\sqrt{2}}\left(\hat{\bm{\theta}} - \mathrm{i} \hat{\bm{\varphi}}\right)\,,& m = -1\,, \\
              \hat{\mathbf{r}}\,,& m = 0\,,
             \\
             -\frac{1}{\sqrt{2}} \left(\hat{\bm{\theta}} + \mathrm{i} \hat{\bm{\varphi}}\right)\,,& m = 1\,.
         \end{cases}$ \\
         \hline
         \text{Cylindrical}& $\Psi^{(1)}_{k_z,0} = 1$& $\mathbf{L}^{(1)}_{k_z,m} = \begin{cases}
             \frac{1}{2} \frac{k_{\rho}}{k} \left( \hat{\bm{\rho}} \pm  \mathrm{i} \hat{\bm{\varphi}} \right)\,,& m = \pm 1\,, \\
             \mathrm{i}\dfrac{k_z}{k}\hat{\mathbf{z}}\,,& m = 0\,.
         \end{cases}$\\
         \hline
    \end{tabular}
    \caption{Regular spherical and cylindrical waves that are nonzero at $\mathbf{r} = \mathbf{0}$.}
    \label{tab:asymptotics}
\end{table}

The total acoustic pressure for a lattice with $N$ scatterers in the unit cell is given by Eq.~\eqref{eq_psca_lattice}. Using Table~\ref{tab:asymptotics}, the corresponding scattered velocity field can be expressed in the SSW basis as
\begin{align}
          \mathbf{v}_{\mathrm{sca}}(\omega; \mathbf{r}) &= -\frac{\rm i}{\rho_{\rm b} c_{\rm b}} \sum_{i = 1}^N \sum_{l,m} \sum_{m'=-1}^1 a^{\rm eff}_{l,m} \Sigma_{i,l,m,1,m'}(\omega, \mathbf{k}_{\parallel}; \mathbf{r}) \mathbf{L}^{(1)}_{1,m'}(\omega;\mathbf{0})\,,
\end{align}
and in the SCW basis as 
\begin{align}
          \mathbf{v}_{\mathrm{sca}}(\omega; \mathbf{r}) &= -\frac{\rm i}{\rho_{\rm b} c_{\rm b}} \sum_{i = 1}^N \sum_{k_z,m}\sum_{m'=-1}^1 a^{\rm eff}_{k_z,m}  \Sigma_{i,k_z,m,k_z,m'}(\omega, \mathbf{k}_{\parallel}; \mathbf{r}) \mathbf{L}^{(1)}_{k_z,m'}(\omega;\mathbf{0})\,.
\end{align}

\section{Expansions of periodic singular waves in another basis}\label{sec_lattice_sums}
To expand periodic singular SSWs in another basis, we first express them as integrals over the Fourier-space spectrum, i.e., in terms of SPWs~\cite{Wittmann1988Aug}:
\begin{align}
\label{eq_sw_integral}
    \begin{aligned}
        \Psi^{(3)}_{l, m}(\omega; \mathbf{r}) = \frac{1}{2 \pi \mathrm{i}^{l}} \iint\limits_{\mathbb{R}^2} \frac{\mathrm{d}k_x \mathrm{d}k_y}{k^2 \gamma_{xy}} Y_{l, m}(\theta_{\mathbf{k}},\varphi_{\mathbf{k}})\mathrm{e}^{\mathrm{i}\mathbf{k}\cdot\mathbf{r}}= \frac{1}{2 \pi \mathrm{i}^{l}} \iint\limits_{\mathbb{R}^2}\frac{\mathrm{d}k_x \mathrm{d}k_z}{k^2 \gamma_{xz}} Y_{l, m}(\theta_{\mathbf{k}},\varphi_{\mathbf{k}})\mathrm{e}^{\mathrm{i}\mathbf{k}\cdot\mathbf{r}}\,,
    \end{aligned}
\end{align}
where $\gamma_{ij} = \sqrt{1 - \frac{k_i^2 + k_j^2}{k^2}}$. In the first representation, $k_z = \pm k \gamma_{xy}$ for $z \gtrless 0$, while in the second, $k_y = \pm k \gamma_{xz}$ for $y \gtrless 0$, reflecting branch cuts in the planes $z = 0$ and $y = 0$, respectively. A similar representation for SCWs is given by~\cite{Cincotti1993Jan}
\begin{align}
    \Psi^{(3)}_{k_z, m}(\omega; \mathbf{r}) = \frac{1}{\pi \mathrm{i}^{m}} \int\limits_{-\infty}^{+\infty} \frac{\mathrm{d}k_x}{k \gamma_{xz}}\mathrm{e}^{\mathrm{i}\mathbf{k}\cdot\mathbf{r}+ \mathrm{i} m \varphi_{\mathbf{k}}}\,,
\end{align}
where $k_y$ is defined as in Eq.~\eqref{eq_sw_integral}. We also employ the Poisson summation formula, which reads here as
\begin{align}
\label{eq_poisson}
    \sum_{\mathbf{R} \in \Lambda_s} \mathrm{e}^{\mathrm{i} \mathbf{k}_{\parallel} \mathbf{R}} = \frac{(2\pi)^d}{V_d}\sum_{\mathbf{G} \in \Lambda_s} \delta^{(d)}(\mathbf{k}_{\parallel} - \mathbf{G})\,,
\end{align}
where $V_d$ is the volume of the unit cell of the lattice $\Lambda_s$, and $\delta^{(d)}(\mathbf{k}_{\parallel} - \mathbf{G})$ is the $d$-dimensional Dirac delta function. 

For a sum of SSWs over a 2D lattice in the $xy$-plane, combining Eqs.~\eqref{eq_sw_integral} and~\eqref{eq_poisson} yields
\begin{align}
    \sum_{\mathbf{R} \in \Lambda_2} \Psi^{(3)}_{l, m}(\omega; \mathbf{r} - \mathbf{R})\mathrm{e}^{\mathrm{i} \mathbf{k}_{\parallel} \mathbf{R}} = \frac{2 \pi L_{l, m}}{A k^2 \mathrm{i}^{l}} \sum_{\mathbf{G} \in \Lambda^{\ast}_2} P^m_{l}(\cos \theta_{\mathbf{k}}) \frac{\mathrm{e}^{\mathrm{i}\mathbf{k}\cdot \mathbf{r} + \mathrm{i}m \varphi_{\mathbf{k}}}}{\sqrt{1 - \frac{(\mathbf{k}_{\parallel} + \mathbf{G})^2}{k^2}}}\,,
\end{align}
where $\mathbf{k} = \mathbf{k}_{\parallel} + \mathbf{G} \pm \sqrt{k^2 - (\mathbf{k}_{\parallel} + \mathbf{G})^2}\hat{\mathbf{z}}$. Using the definition of SPWs in Eq.~\eqref{eq_spw_def}, one obtains Eq.~\eqref{eq_sw_into_pw_lattice}.

\section{Acoustic \textit{T}-matrix of a multilayered cylinder}\label{sec_cylinder}
Within \textit{acoustotreams}, the acoustic \textit{T}-matrix of an infinite cylinder can be computed for three types of interface conditions: fluid-fluid, hard-fluid, and solid-fluid. In the fluid-fluid case, the cylinder may consist of multiple concentric layers defined by radii $R_i$ and material parameters $(\rho_i, c_i)$, with $i = 1, \ldots, \mathcal{N}$. Formally, $(\rho_{\rm b}, c_{\rm b}) \equiv (\rho_{\mathcal{N}+1}, c_{\mathcal{N}+1})$.

To obtain the \textit{T}-matrix of the $(i+1)$th layer from that of the $i$th layer, we follow Ref.~\cite{Sainidou2005Mar} and apply the interface conditions at the interface between layers $i$ and $(i+1)$, namely the continuity of pressure and the normal component of velocity. Such a procedure leads to the system
\begin{align}
    \begin{pmatrix}
    J_m(k_{\rho}^i R_i) + T_{m}^i H_m^{(1)}(k_{\rho}^i R_i)& -H_m^{(1)}(k^{i+1}_{\rho}R_{i+1}) \\
        \delta \left[J'_m(k_{\rho}^i R_i) + T_{m}^i H^{(1)\prime}_m(k_{\rho}^i R_i)\right]& -H_m^{(1)\prime}(k^{i+1}_{\rho}R_{i+1})  
    \end{pmatrix}
    \begin{pmatrix}
        T_{m}^{i+1} \\
        M_{m}^{i+1} 
    \end{pmatrix}
    =
    \begin{pmatrix}
    J_m(k^{i+1}_{\rho}R_{i+1}) \\
        J'_m(k^{i+1}_{\rho}R_{i+1})
    \end{pmatrix}\,,
\end{align}
where $k_{\rho}^i = \sqrt{(\omega/c_i)^2 - k_z^2}$, $\delta = \frac{\rho_{i + 1}k_{\rho}^i}{\rho_ik_{\rho}^{i+1}}$. Here, $T_{m}^{1} \equiv 0$, and the desired \textit{T}-matrix element is $T_{k_z,m;k_z,m} \equiv T_{m}^{\mathcal{N}+1}$, while $M_{mm}^{i+1}$ relates the incident and internal field coefficients. 

For hard and soft cylindrical interfaces, the \textit{T}-matrix elements reduce to~\cite{Roumeliotis2001Mar}
\begin{align}
    T_{k_z,m;k_z,m} = -\frac{J_m(k_{\rho}^{\rm b} R)}{H^{(1)}_m(k_{\rho}^{\rm b} R)}\,, \quad T_{k_z,m;k_z,m} = -\frac{J'_m(k_{\rho}^{\rm b} R)}{H^{(1)\prime}_m(k_{\rho}^{\rm b} R)}\,,
\end{align}
respectively.

\section{Physical constraints on the acoustic \textit{T}-matrix in the SSW basis}\label{sec_comsol}
The acoustic \textit{T}-matrix of a scatterer must satisfy several physical constraints imposed by its material and geometrical properties~\cite{Asadova2025Mar}.

\subsection{Nonabsorptive scatterer}
For a nonabsorptive scatterer, energy conservation implies (see Ref.~\cite{Waterman2009Jan}, or Eq.~(7.68) in Ref.~\cite{Martin2006Aug}):
\begin{align}
    \mathbf{T}\mathbf{T}^\dagger = -\frac{1}{2} \left( \mathbf{T} + \mathbf{T}^\dagger\right)\,.
\end{align}

\subsection{Passive scatterer}
For a passive scatterer (dissipation exceeds gain), the operator
\begin{align}
    -2\mathbf{T}\mathbf{T}^\dagger - \left( \mathbf{T} + \mathbf{T}^\dagger\right)\,,
\end{align}
must be Hermitian positive semidefinite~\cite{LeRu2013Jan}.

\subsection{Reciprocity}
The reciprocity theorem in acoustics~\cite{Pierce2019} implies (Eq.~(7.67) in Ref.~\cite{Martin2006Aug})
\begin{align}
    T_{l,m,l',m'} = (-1)^{m+m'} T_{l',-m',l,-m}\,.
\end{align}

\subsection{Rotational symmetry}
For scatterers with ${\mathfrak{n}}$-fold rotational symmetry about the $z$-axis, invariance under rotation by $2\pi / \mathfrak{n}$, together with the property,
\begin{align}
    Y_{l,m}(\pi,\varphi + 2\pi / \mathfrak{n}) = \mathrm{e}^{\mathrm{i} 2\pi m/ \mathfrak{n}} Y_{l,m} (\theta,\varphi)\,,
\end{align}
implies
\begin{align}
    T_{l,m,l',m'} = \delta_{m,m'+s\mathfrak{n}}T_{l,m,l',m'}\,,
\end{align}
where $s$ is an integer. In particular, $m = m'$ for cylindrical rotational symmetry ($\mathfrak{n} = \infty$).

\subsection{Mirror symmetry}
Under reflection $z \to -z$, the spherical harmonics transform as
\begin{align}
    Y_{l,m}(\pi-\theta,\varphi) = (-1)^{l+m} Y_{l,m} (\theta,\varphi)\,.
\end{align}
For scatterers with this symmetry, the \textit{T}-matrix must satisfy 
\begin{align}
    T_{l,m,l',m'} = (-1)^{l+m+l'+m'}T_{l,m,l',m'}\,.
\end{align}

\subsection{Inversion symmetry}
The parity of the spherical harmonics with respect to the coordinate inversion $\mathbf{r} \to -\mathbf{r}$ reads as
\begin{align}
    Y_{l,m}(\pi-\theta,\pi + \varphi) = (-1)^{l} Y_{l,m} (\theta,\varphi)\,.
\end{align}
For centrosymmetric scatterers, parity conservation implies
\begin{align}
    T_{l,m,l',m'} = \delta_{l,l' + 2s}T_{l,m,l',m'}\,,
\end{align}
where $s$ is a nonnegative integer.





\bibliographystyle{elsarticle-num}
\bibliography{main}







\end{document}